\documentclass[aps,floatfix,prx,superscriptaddress,reprint,showpacs,10pt,preprintnumbers,longbibliography]{revtex4-2}
\usepackage[utf8]{inputenc}
\usepackage[pdftex]{graphicx}
\usepackage{float}
\usepackage{amssymb}
\usepackage{amsmath}  
\usepackage{amsfonts}
\usepackage{amsthm}

\usepackage{dsfont}
\usepackage{array}
\usepackage{bm}
\usepackage{mathrsfs}
\usepackage{pifont}
\usepackage{multirow}
\usepackage{upgreek}
\usepackage[dvipsnames]{xcolor}
\usepackage[pdftex,
            pdftitle={Quantum Computing for High-Energy Physics},
            pdfauthor={Authors},
            bookmarks,
            colorlinks,
            linkcolor=black,
            citecolor=mymagenta,
            menucolor=black,
            urlcolor=myblue,
            plainpages=false,
            pdfpagelabels,
            hypertexnames=false]{hyperref}
\usepackage{verbatim}
\usepackage{slashed}
\usepackage{bm}
\usepackage{braket}
\usepackage{booktabs}
\usepackage{multirow}
\usepackage{physics}
\usepackage{bm}
\usepackage{bbm}
\usepackage{tabularx}
\usepackage{longtable}
\usepackage{textcomp}
\usepackage[acronym,symbols,nogroupskip,nomain,nonumberlist,nopostdot,toc]{glossaries}
\definecolor{mymagenta}{RGB}{200, 0, 100}
\definecolor{myblue}{RGB}{45, 48, 146}

\graphicspath{{figures/}}

\newacronym{hep}{HEP}{High-Energy Physics}
\newacronym{qc}{QC}{Quantum Computing}
\newacronym{lgt}{LGT}{Lattice Gauge Theory}
\newacronym{qed}{QED}{Quantum Electrodynamics}
\newacronym{qcd}{QCD}{Quantum Chromodynamics}
\newacronym{tn}{TN}{Tensor Network}
\newacronym{qtn}{QTN}{Quantum Tensor Network}
\newacronym{vqe}{VQE}{Variational Quantum Eigensolver}
\newacronym{vqd}{VQD}{Variational Quantum Deflation}
\newacronym{ssvqe}{SSVQE}{Subspace-search Variational Quantum Eigensolver}
\newacronym{pf}{PF}{Product Formulas}
\newacronym{vte}{VTE}{Variational Time Evolution}
\newacronym{eom}{EOM}{Equation of Motion}
\newacronym{1p1D}{(1+1)D}{(1+1)-Dimensional}
\newacronym{2p1D}{(2+1)D}{(2+1)-Dimensional}
\newacronym{3p1D}{(3+1)D}{(3+1)-Dimensional}
\newacronym{cnot}{CNOT}{Controlled NOT gate}
\newacronym{mps}{MPS}{Matrix Product States}
\newacronym{zne}{ZNE}{Zero-Noise Extrapolation}
\newacronym{pec}{PEC}{probabilistic error cancellation}
\newacronym{qgan}{QGAN}{Quantum Generative Adversarial Network}
\newacronym{aqgan}{QGAN}{Anomaly Quantum Generative Adversarial Network}
\newacronym{qubo}{QUBO}{Quantum Unconstrained Binary Optimization}
\newacronym{hhl}{HHL}{Harrow-Hassidim-Lloyd}
\newacronym{qnn}{QNN}{Quantum Neural Network}
\newacronym{qaoa}{QAOA}{Quantum Approximate Optimization Algorithm}
\newacronym{qsvm}{QSVM}{Quantum-enhanced Support Vector Machine}
\newacronym{mcmc}{MCMC}{Markov Chain Monte Carlo}
\newacronym{bsm}{BSM}{Beyond the Standard Model}
\newacronym{sm}{SM}{Standard Model}
\newacronym{ks}{K-S}{Kogut-Susskind}
\newacronym{qkmeans}{QKMEANS}{Quantum k-means}
\newacronym{tda}{TDA}{Topological Data Analysis}
\newacronym{rl}{RL}{Reinforcement Learning}
\newacronym{qrl}{QRL}{Quantum Reinforcement Learning}
\newacronym{gnn}{GNN}{Graph Neural Network}
\newacronym{qgnn}{QGNN}{quantum-classical hybrid Graph Neural Network}
\newacronym{gpu}{GPU}{Graphics Processing Unit}
\newacronym{qml}{QML}{Quantum Machine Learning}
\newacronym{gqml}{GQML}{Geometrical Quantum Machine Learning}
\newacronym{vqa}{VQA}{Variational Quantum Algorithm}
\newacronym{qae}{QAE}{Quantum AutoEncoder}
\newacronym{csvc}{CSVC}{Classic Support Vector Classifier}
\newacronym{qsvc}{QSVC}{Quantum Support Vector Classifier}
\newacronym{nisq}{NISQ}{Noisy Intermediate-Scale Quantum}
\newacronym{vbs}{VBS}{Vector Boson Scattering}
\newacronym{ctf}{CTF}{Combinatorial Track Finder}
\newacronym{irc}{IRC}{Infrared and Collinear}
\newacronym{pdf}{PDF}{Parton Distribution Function}
\newacronym{eft}{EFT}{Effective Field Theory}
\newacronym{pdfs}{PDFs}{Parton Distribution Functions}
\newacronym{efts}{EFTs}{Effective Field Theories}
\newacronym{pqc}{PQC}{Parameterized Quantum Circuit}
\newacronym{varQTE}{varQTE}{variational Quantum Time Evolution}
\newacronym{varQITE}{varQITE}{variational Quantum Imaginary Time Evolution}
\newacronym{qcnn}{QCNN}{Quantum Convolutional Neural Network}
\newacronym{qcbm}{QCBM}{Quantum Circuit Born Machine}
\newacronym{povm}{POVM}{Positive Operator Valued Measure}
\newacronym{msw}{MSW}{Mikheyev-Smirnov-Wolfenstein}
\newacronym{cms}{CMS}{Compact Muon Solenoid}
\newacronym{qbm}{QBM}{Quantum Boltzmann Machine}

\makeglossaries

\newcommand{\oper}[1]{\hat{#1}} 

\newcommand{\Z}{{\mathbb{Z}}}

\newcommand{\CP}{{\mathbb{C}P}}
\newcommand{\1}{1\!\!1}

\newcommand{\p}{\partial}

\begin{document}
\title{Quantum Computing for High-Energy Physics \\ State of the Art and Challenges \\ Summary of the QC4HEP Working Group}

\author{Alberto~Di~Meglio}
\email{alberto.di.meglio@cern.ch}
\affiliation{European Organization for Nuclear Research (CERN), CH-1211 Geneva, Switzerland}

\author{Karl~Jansen}
\email{karl.jansen@desy.de}
\affiliation{CQTA, Deutsches Elektronen-Synchrotron DESY, Platanenallee 6, 15738 Zeuthen, Germany}
\affiliation{Computation-based Science and Technology Research Center, The Cyprus Institute, 20, Constantinou Kavafi str., 2121 Nicosia, Cyprus}

\author{Ivano~Tavernelli}
\email{ita@zurich.ibm.com}
\affiliation{IBM Quantum, IBM Research – Zurich, 8803 R\"uschlikon, Switzerland}

\author{Constantia~Alexandrou}
\affiliation{Department of Physics, University of Cyprus, PO Box 20537, 1678 Nicosia, Cyprus}
\affiliation{Computation-based Science and Technology Research Center, The Cyprus Institute, 20, Constantinou Kavafi str., 2121 Nicosia, Cyprus}

\author{Srinivasan~Arunachalam}
\affiliation{IBM Quantum, IBM Research - 1101 Kitchawan Rd, Yorktown Heights, NY, USA}

\author{Christian~W.~Bauer}
\affiliation{Physics Division LBNL - M/S 50A5104 1 Cyclotron Rd Berkeley, CA,  USA}

\author{Kerstin~Borras}
\affiliation{Deutsches Elektronen-Synchrotron DESY, Notkestrasse 85, 22607 Hamburg, Germany}
\affiliation{RWTH Aachen University, Templergraben 55, 52062 Aachen, Germany}

\author{Stefano~Carrazza}
\affiliation{TIF Lab, Dipartimento di Fisica, Università degli Studi di Milano and INFN Sezione di Milano, Milan, Italy}
\affiliation{European Organization for Nuclear Research (CERN), CH-1211 Geneva, Switzerland}

\author{Arianna~Crippa}
\affiliation{CQTA, Deutsches Elektronen-Synchrotron DESY, Platanenallee 6, 15738 Zeuthen, Germany}
\affiliation{Institut für Physik, Humboldt-Universit\"at zu Berlin, Newtonstr. 15, 12489 Berlin, Germany}

\author{Vincent~Croft}
\affiliation{$\langle aQa^L\rangle$ Applied Quantum Algorithms Leiden, The Netherlands}

\author{Roland~de~Putter}
\affiliation{IBM Quantum, IBM Research - 1101 Kitchawan Rd, Yorktown Heights, NY, USA}

\author{Andrea~Delgado}
\affiliation{Physics Division, Oak Ridge National Laboratory, Oak Ridge, TN, 37831, USA}

\author{Vedran~Dunjko}
\affiliation{$\langle aQa^L\rangle$ Applied Quantum Algorithms Leiden, The Netherlands}

\author{Daniel~J.~Egger}
\affiliation{IBM Quantum, IBM Research – Zurich, 8803 R\"uschlikon, Switzerland}

\author{Elias~Fernández-Combarro}
\affiliation{Department of Computer Science, Facultad de Ciencias, University of Oviedo, 33007, Asturias, Spain }

\author{Elina~Fuchs}
\affiliation{European Organization for Nuclear Research (CERN), CH-1211 Geneva, Switzerland}
\affiliation{Institute of Theoretical Phyiscs, Leibniz University Hannover, 30167 Hannover, Germany}
\affiliation{Physikalisch-Technische Bundesanstalt, 38116 Braunschweig, Germany}

\author{Lena~Funcke}
\affiliation{Transdisciplinary Research Area ``Building Blocks of Matter and Fundamental Interactions'' (TRA Matter) and Helmholtz Institute for Radiation and Nuclear Physics (HISKP), University of Bonn, Nußallee 14-16, 53115 Bonn, Germany}

\author{Daniel~Gonz\'alez-Cuadra}
\affiliation{Institute for Theoretical Physics, University of Innsbruck, 6020 Innsbruck, Austria}
\affiliation{Institute for Quantum Optics and Quantum Information of the Austrian Academy of Sciences,
6020 Innsbruck, Austria}

\author{Michele~Grossi}
\affiliation{European Organization for Nuclear Research (CERN), CH-1211 Geneva, Switzerland}

\author{Jad~C.~Halimeh}
\affiliation{Department of Physics and Arnold Sommerfeld Center for Theoretical Physics, Ludwig-Maximilians-Universit\"at M\"unchen,Germany}
\affiliation{Munich Center for Quantum Science and Technology, Germany}

\author{Zo\"{e}~Holmes}
\affiliation{Institute of Physics, Ecole Polytechnique F\'{e}d\'{e}rale de Lausanne (EPFL), CH-1015 Lausanne, Switzerland}

\author{Stefan~Kühn}
\affiliation{CQTA, Deutsches Elektronen-Synchrotron DESY, Platanenallee 6, 15738 Zeuthen, Germany}

\author{Denis~Lacroix}
\affiliation{Paris-Saclay University, CNRS/IN2P3, IJCLab, 91405 Orsay, France}

\author{Randy~Lewis}
\affiliation{Department of Physics and Astronomy, York University, Toronto, Ontario, Canada, M3J 1P3}

\author{Donatella~Lucchesi}
\affiliation{Department of Physics and Astronomy, Università di Padova,V. Marzolo 8, I-35131 Padova, Italy}
\affiliation{INFN - Sezione di Padova, Via Marzolo 8, 35131 Padova, Italy}
\affiliation{European Organization for Nuclear Research (CERN), CH-1211 Geneva, Switzerland}

\author{Miriam~Lucio~Martinez}
\affiliation{Nikhef – National Institute for Subatomic Physics, Science Park 105, 1098 XG Amsterdam, The Netherlands}
\affiliation{Department of Gravitational Waves and Fundamental Physics, Maastricht University, 6200 MD Maastricht, The Netherlands}

\author{Federico~Meloni}
\affiliation{Deutsches Elektronen-Synchrotron DESY, Notkestrasse 85, 22607 Hamburg, Germany}

\author{Antonio~Mezzacapo}
\affiliation{IBM Quantum, IBM Research - 1101 Kitchawan Rd, Yorktown Heights, NY, USA}

\author{Simone~Montangero}
\affiliation{Department of Physics and Astronomy, Università di Padova,V. Marzolo 8, I-35131 Padova, Italy}
\affiliation{INFN - Sezione di Padova, Via Marzolo 8, 35131 Padova, Italy}

\author{Lento~Nagano}
\affiliation{International Center for Elementary Particle Physics (ICEPP), The University of Tokyo, 7-3-1 Hongo, Bunkyo-ku, Tokyo 113-0033, Japan}

\author{Voica~Radescu}
\affiliation{IBM Quantum, IBM Deutschland Research \& Development GmbH - Schoenaicher Str. 220, 71032 Boeblingen, Germany}

\author{Enrique~Rico~Ortega}
\affiliation{Department of Physical Chemistry, University of the Basque Country UPV/EHU, Box 644, 48080 Bilbao, Spain}
\affiliation{Donostia International Physics Center, 20018 Donostia-San Sebastián, Spain}
\affiliation{EHU Quantum Center, University of the Basque Country UPV/EHU, P.O. Box 644, 48080 Bilbao, Spain}
\affiliation{IKERBASQUE, Basque Foundation for Science, Plaza Euskadi 5, 48009 Bilbao, Spain}

\author{Alessandro~Roggero}
\affiliation{Department of Physics, University of Trento, via Sommarive 14, I–38123, Povo, Trento, Italy}
\affiliation{INFN-TIFPA Trento Institute of Fundamental Physics and Applications, via Sommarive 14, I–38123, Trento, Italy}

\author{Julian~Schuhmacher}
\affiliation{IBM Quantum, IBM Research – Zurich, 8803 R\"uschlikon, Switzerland}

\author{Joao~Seixas}
\affiliation{Instituto Superior Técnico, Dep. Física, Lisboa, Portugal}
\affiliation{Center of Physics and Engineering of Advanced Materials (CeFEMA), Instituto Superior Técnico, Lisboa, Portugal,}
\affiliation{Laboratory of Physics for Materials and Emergent Technologies (LaPMET), Portugal}

\author{Pietro~Silvi}
\affiliation{Department of Physics and Astronomy, Università di Padova,V. Marzolo 8, I-35131 Padova, Italy}
\affiliation{INFN - Sezione di Padova, Via Marzolo 8, 35131 Padova, Italy}

\author{Panagiotis~Spentzouris}
\affiliation{Fermi National Accelerator Laboratory, Kirk and, Pine St, Batavia, IL 60510, USA}

\author{Francesco~Tacchino}
\affiliation{IBM Quantum, IBM Research – Zurich, 8803 R\"uschlikon, Switzerland}

\author{Kristan~Temme}
\affiliation{IBM Quantum, IBM Research - 1101 Kitchawan Rd, Yorktown Heights, NY, USA}

\author{Koji~Terashi}
\affiliation{International Center for Elementary Particle Physics (ICEPP), The University of Tokyo, 7-3-1 Hongo, Bunkyo-ku, Tokyo 113-0033, Japan}

\author{Jordi~Tura}
\affiliation{$\langle aQa^L\rangle$ Applied Quantum Algorithms Leiden, The Netherlands}
\affiliation{Instituut-Lorentz, Universiteit Leiden, P.O. Box 9506, 2300 RA Leiden, The Netherlands}

\author{Cenk~Tüysüz}
\affiliation{CQTA, Deutsches Elektronen-Synchrotron DESY, Platanenallee 6, 15738 Zeuthen, Germany}
\affiliation{Institut für Physik, Humboldt-Universit\"at zu Berlin, Newtonstr. 15, 12489 Berlin, Germany}

\author{Sofia~Vallecorsa}
\affiliation{European Organization for Nuclear Research (CERN), CH-1211 Geneva, Switzerland}

\author{Uwe-Jens~Wiese}
\affiliation{Albert Einstein Center for Fundamental Physics, Institute for Theoretical Physics, University of Bern, Sidlerstrasse 5, CH-3012 Bern, Switzerland}

\author{Shinjae~Yoo}
\affiliation{Brookhaven National Laboratory, 98 Rochester St, Upton, NY 11973, USA}

\author{Jinglei~Zhang}
\affiliation{Institute for Quantum Computing, University of Waterloo, Waterloo, ON, Canada, N2L 3G1}
\affiliation{Department of Physics \& Astronomy, University of Waterloo, Waterloo, ON, Canada, N2L 3G1}

\date{\today}

\begin{abstract}
  Quantum computers offer an intriguing path for a paradigmatic change of computing in the natural sciences and beyond, with the potential for achieving a so-called quantum advantage, namely a significant (in some cases exponential) speed-up of numerical simulations. The rapid development of hardware devices with various realizations of qubits enables the execution of small scale but representative applications on quantum computers. In particular, the high-energy physics community plays a pivotal role in accessing the power of quantum computing, since the field is a driving source for challenging computational problems. This concerns, on the theoretical side, the exploration of models which are very hard or even impossible to address with classical techniques and, on the experimental side, the enormous data challenge of newly emerging experiments, such as the upgrade of the Large Hadron Collider. In this roadmap paper, led by CERN, DESY and IBM, we provide the status of high-energy physics quantum computations and give examples for theoretical and experimental target benchmark applications, which can be addressed in the near future. Having the IBM $100 \otimes 100$ challenge in mind, where possible, we also provide resource estimates for the examples given using error mitigated quantum computing.  
\end{abstract}

\maketitle



\section{Introduction}
\label{Introduction}

This article reports on scientific discussions and conclusions elaborated at a workshop on \gls{hep} held in November 2022 at CERN in Geneva.
This first event of the Quantum Computing for HEP (QC4HEP) Working Group gathered experts on \gls{hep} from different academic and research institutions and countries over four continents, who besides being world experts in theoretical and experimental aspects of \gls{hep}, also shared a common interest in \gls{qc} and its potential as a game changer in the field. 
The main goal of the workshop, and of this report-article, is to set a common roadmap for selected topics of interest to this community, in which we believe that \gls{qc} can have a significant impact in the near future. 
To this end, we have investigated classes of problems and corresponding quantum algorithms that can lead to potential quantum advantage with near-term, noisy, quantum devices, and - in particular - using IBM superconducting devices. We aim at delivering a set of physically relevant use cases that can become interesting 
demonstrations in view of the $100 \otimes 100$ challenge announced by IBM~\cite{IBM_100by100}. 

For practical purposes, we have organized this article into two main domain areas: theoretical methods and algorithms for modelling \gls{hep} problems, and numerical methods for the interpretation and analysis of experimental results as well as detector simulation and event generation. 
We strongly believe that there are important connections between the two research domains, where many of the quantum algorithms designed for the solution of problems in one field can also be transferred to the other. 

We will therefore start with a short summary of the main \gls{hep} domains in theoretical modelling and experimental physics, for which we believe there is the potential for quantum computing to play a significant role in the near-term.

\subsection{Quantum Computing for Theoretical Modelling in \gls{hep}}

Despite the great success of classical lattice field theory (e.g., for \gls{qed} and \gls{qcd} simulations~\cite{Durr2008,Alexandrou2020}), out-of-equilibrium and real-time dynamics (e.g., of particle collisions, thermalization phenomena or dynamics after a quench), remain out of reach for euclidean path-integral Monte Carlo simulations. 
Furthermore, properties of nuclear matter at high fermionic densities, as they arise in neutron stars or at the very early universe  for example, can not be accessed through these classical simulation techniques~\cite{Fukushima2011}. The same holds true for theories with topological terms, which are relevant, e.g. in \gls{qcd} for understanding the amount of CP-violation or, in the electroweak sector, the sphaleron rate in the early universe.  
These severe limitations are rooted in the notorious \emph{sign-problem}: the highly oscillatory behaviour of the path integrals arising in real-time phenomena, in systems with a high fermionic particle density or in the presence of topological terms imply an exponentially growing sampling run-time complexity with an increasing number of lattice sites~\cite{Troyer2005}.

An alternative approach to circumvent the sign problem might be to describe lattice fields theories in the equivalent Hamiltonian formalism, instead of the path integral description based on the Lagrangian formalism~\cite{Kogut1975,Kogut1979}. 
In the Hamiltonian approach, however, the total many-particle wave function which describes a general particle state on the whole lattice must be stored throughout the simulation. 
But since the total discretized Hilbert space $\mathcal{H}$ containing these general states corresponds to a tensor product of Hilbert spaces $\mathcal{H}_j$ on a single lattice site, the required memory to store a full wave function on the lattice scales exponentially with the number of lattice sites.

In recent years, novel tensor network-based methods have been introduced to alleviate these limitations by allowing for a more compact representation of general quantum states on the lattice~\cite{Silvi2014,Dalmonte2016,Banuls2019SimulatingLG,Banuls2019,Banuls2020TNreview}. The underlying mechanism which allows Hamiltonian simulations to be performed is that only a small subspace of the complete Hilbert space describes the low energy physics of quantum field theories and \gls{tn} methods identify and focus exactly on these physically relevant subspaces. Hence, with tensor network techniques, various phenomena such as string breaking and real-time dynamics~\cite{Buyens2013,Kuehn2015,Pichler2016,Buyens2016b,Banuls2019b,Rigobello2021} or phase diagrams of both abelian and non-abelian gauge theories at finite fermionic densities~\cite{Banuls2016a,Silvi2017,Felser2019,Silvi2019} have been studied on a few hundred lattice sites at least in one space dimensional models. 

A very promising alternative to \gls{tn} are simulations on quantum computers which can represent large Hilbert spaces using qubits, its basic unit of information, where the number of required qubits merely grows linearly with the number of lattice sites. Moreover, quantum algorithms have been proposed that implement real-time dynamics with polynomial time complexity for scalar quantum field theories and \gls{qed}~\cite{Byrnes2006, Jordan2012, Mathis2020}. 
In addition, by sharing with tensor networks the Hamiltonian formulation, quantum computations completely avoid the sign problem. 
Thus, quantum computers offer a potential framework to fully overcome the limitations outlined above for the simulation of lattice gauge theories and especially their real-time dynamics~\cite{Feynman1982}. 

Indeed, various proposals for the implementation of general abelian and non-abelian \gls{lgt} on different types of quantum hardware have accumulated in the past few years, and simulations of small \gls{lgt} systems on real quantum devices have been demonstrated~\cite{Banuls2019SimulatingLG,Banuls2019,Klco2019,Atas2021,Ciavarella2021,Clemente2022a}. Examples include proposals for implementing lattice gauge theories using optical lattices~\cite{Banerjee2012,Tagliacozzo2013,Tagliacozzo2013a}, atomic and ultra-cold quantum matter~\cite{Buchler2005,Zohar2011,Zohar2012,Hauke2013,Zohar2013a,Zohar2013c,Banerjee2013,Zohar2015a,Laflamme2015,Gonzalez-Cuadra2017,Rico2018,Zache2018}, further proof-of-principle implementations on a real superconducting architecture~\cite{Klco2018,Klco2019,Atas2021,Ciavarella2021,Mazzola2021} and ultimately, \gls{1p1D} real-time and variational simulations of quantum electrodynamics on a trapped ion system~\cite{Martinez2016, Kokail2019}. A broad overview of recently proposed quantum simulators and implementation techniques for \gls{lgt} can be found in~\cite{Dalmonte2016,Banuls2019SimulatingLG,Banuls2019}. It is noteworthy that lattice gauge theories can be approached by many different physical systems and methods, each featuring its own advantages and disadvantages.

The understanding of the static and dynamical properties of \gls{3p1D} \gls{lgt}, including \gls{qed} and \gls{qcd}, is not the only target of today's theoretical particle physics. In fact, one has to consider an exciting but also demanding roadmap to reach eventually the goal of performing quantum simulations of \gls{3p1D} systems as relevant for \gls{hep}. This roadmap starts with \gls{1p1D} systems which are under active research nowadays, moving to \gls{2p1D} systems which are under consideration already now by various groups and reach \gls{3p1D} systems in the future. 

Lower dimensional systems in \gls{1p1D} and \gls{2p1D} dimensions are already very interesting. They share important and challenging problems with their higher-dimensional counterparts. One important example is the study of \gls{2p1D} \gls{qed} which shows the phenomena of asymptotic freedom and confinement. Asymptotic freedom is a feature of \gls{qcd}, i.e., the quantum field theory of the strong interaction between quarks and gluons. In the limit of high energies (small distances when natural units are used) the quarks become weakly interacting making perturbation theory well suited for theoretical predictions. On the contrary, at low energies the interaction becomes strong leading to particle confinement. Interestingly enough, there are also low dimensional \gls{lgt} for which the phenomena of confinement is known, which can help  shedding new lights on the theoretically harder \gls{qcd} confining mechanism (because of the large dimensionality and the high number of degrees of freedom).
As said above, one such model is \gls{2p1D} \gls{qed}, which is a compact $U(1)$ \gls{lgt}. As outlined in Section \ref{subsubsect_2+1QED},  we therefore propose this model in a lower dimension as a benchmark 
for exploring the potential of quantum computing in the near-term, noisy, regime. 

\subsection{Quantum Computing in \gls{hep} Experiments}
\gls{hep} experiments are characterised by the ability to probe the intricacies of particle physics in the Standard Model and beyond it, through performing measurements and analyses at the frontier between quantum theory and precision experimentation.
The statistical precision of experiments performed at elementary particles   scales is predicated on three classes of algorithms:
\begin{itemize}
    \item Detector operation algorithms allow detectors to efficiently obtain data that cleanly represents the fundamental interactions of matter. These detectors might feature very large amounts of very high dimensional data such as those found inside hadron colliders. These detectors require algorithms to sort significant signals from noise. Detector-based algorithms are also used to aid in inferring more complete features of a given measurement of very rare processes such as neutrino or expected New Physics interactions.
    \item Identification and reconstruction algorithms are an essential part of mapping the vast collection of pixel intensities, timings, and event counts to a coherent underlying physics structure in the data. These algorithms allow the segmentation of datasets into those which feature particular processes or states that are relevant to a given physics goal and therefore must be robust, efficient, and unbiased.
    \item Robust simulation and inference tools allow 
    particle physics experiments to compare large amounts of complex, highly structured data with parameterized theoretical predictions. These algorithms include the creation of simulated datasets that are used as templates in parametric statistical models, classification tools to enhance the sensitivity of a given measurement to some process, or the identification of statistically anomalous signals that might hint at sources of new physics. 
\end{itemize}
\gls{qc} 
encompasses
several defining characteristics that are of particular interest to experimental \gls{hep}: the potential for quantum speed-up in processing time, sensitivity to sources of correlations in data, and increased expressivity of quantum systems. 
Each of the three classes of algorithms mentioned above benefits from all three of these characteristics. Experiments running on high-luminosity accelerators need faster algorithms; identification and reconstruction algorithms need to capture correlations in signals; simulation and inference tools need to express and calculate functions that are classically intractable.

Within the existing data reconstruction and analysis paradigm, access to algorithms that exhibit quantum speed-ups would revolutionise the simulation of large-scale quantum systems and the processing of data from complex experimental set-ups. This would enable a new generation of precision measurements to probe deeper into the nature of the universe. Existing measurements may contain the signatures of underlying quantum correlations or other sources of new physics that are inaccessible to classical analysis techniques. Quantum algorithms that leverage these properties could potentially extract more information from a given dataset than classical algorithms. Finally, algorithms that can capture more complex aspects of \gls{hep} theory and simulation could provide estimators that are more natively aligned with the quantum mechanical nature of the Standard Model or indeed potentially uncover new physics beyond what can be explained by classical models.

Quantum computing for \gls{hep} is of particular interest due to the prospect of algorithms that can leverage the unique properties of quantum systems to achieve computational advantages. Most quantum algorithms with a promise of a super-polynomial advantage exploit the capacity of quantum computers to efficiently simulate quantum-many-body systems. The search for potential quantum advantage would be accelerated by the identification of computational problems with the right kind of underlying structure which can be leveraged by quantum algorithms. Applications in the \gls{hep} domain can clearly offer a controlled experimental benchmark for such test cases. Through the analysis of the data from \gls{hep} experiments using quantum algorithms, researchers may be able to gain insights into the behaviour of quantum systems and potentially identify new avenues for quantum advantage.

\gls{hep} experimental data is typically organized as collections of associated detector signals that can be reconstructed into measured particles. The distributions of these particle measurements are calculable under specific parameterization of the underlying theory such that the distribution of experimental data can be directly compared to theoretical predictions through the use of simulated data.
These parameterizations are such that a characterisation of any given process as defined in quantum field theory is maximally described by the data. 
This method of parameterization allows the accuracy of the estimator to scale consistently and efficiently with repeated measurements.
Therefore, although the data recorded in high-energy physics experiments provide information about the behaviour of fundamental particles and their interactions, which in turn are described by quantum fields and their dynamics governed by the principles of quantum mechanics, it is important to note that typically the data and their descriptions are classical in nature and therefore may not trivially exhibit the quantum mechanical properties necessary for quantum advantage. 
In summary, by analyzing experimental data using tools and techniques from both quantum information theory and particle physics, we can gain insights into the fundamental nature of the universe and potentially discover new phenomena that are not yet understood.

It is worth mentioning that another community article on quantum simulations for \gls{hep} appeared recently in the literature~\cite{PRXQuantum.4.027001}. 
Despite the broadly similar target, our work differentiates in several essential aspects; first, our focus is on the identification and detailed characterization of projects that - while approachable with near-term, noisy quantum devices (within the $100 \otimes 100$ challenge) -  can already address problems of interest in the \gls{hep} community. Second, our investigation comprises both theoretical models as well as computational aspects related to particle collision experiments.

\begin{figure}[htp!]
    \centering
    \includegraphics[width=.95 \columnwidth, keepaspectratio]{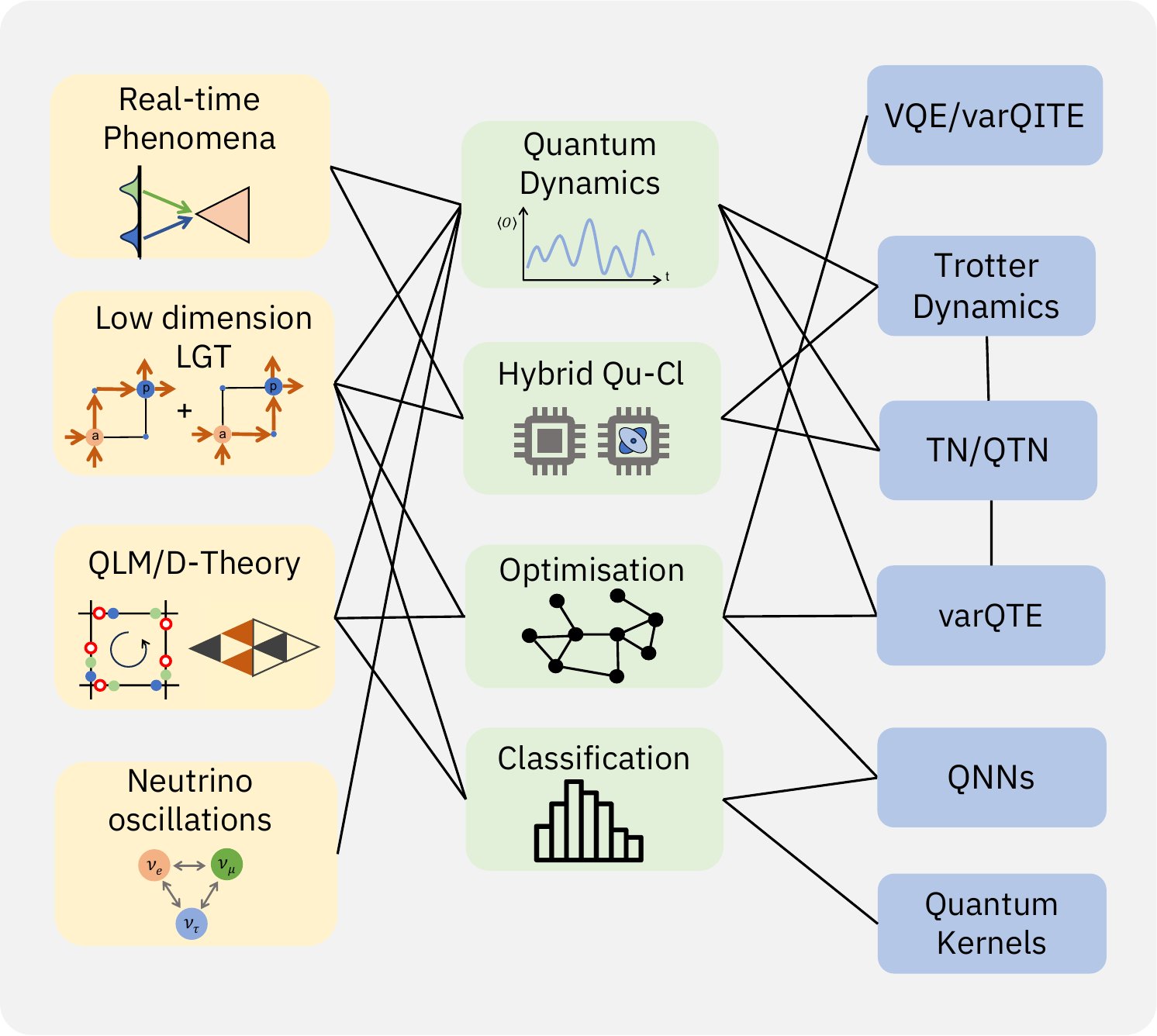} \\
    \includegraphics[width=.95 \columnwidth, keepaspectratio]{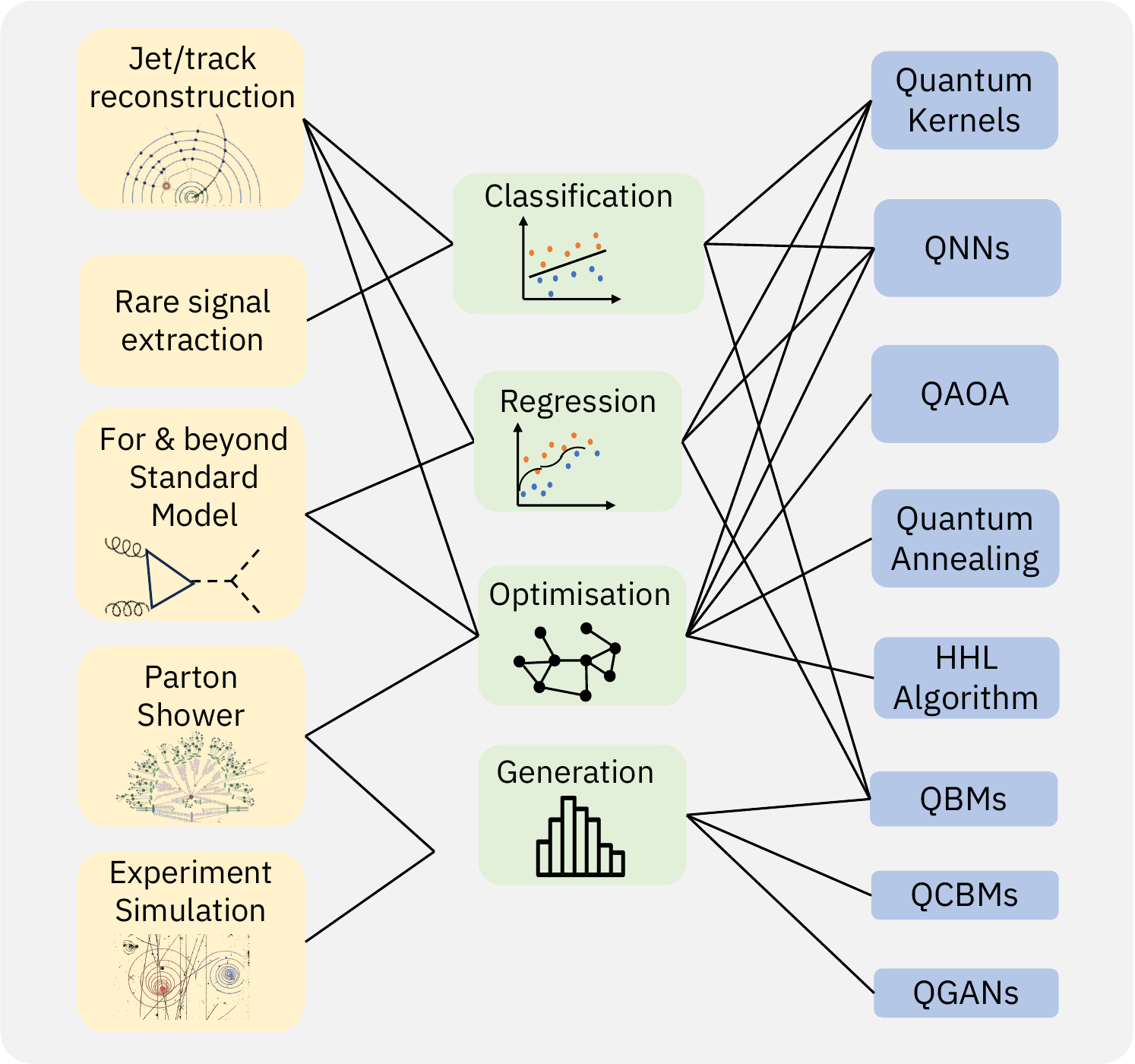}
    \caption{(Upper Panel) Proposed theoretical physical model systems (orange) with corresponding approaches (green) and quantum algorithms (blue). For more information on the identified areas of interest see Section~\ref{subsect_Theory}. 
    (Lower Panel) 
    Proposed experimental challenges (orange) with corresponding approaches (green) and quantum algorithms (blue). For more information on the identified areas of interest see Section~\ref{subsect_Experiments}. 
    Legend: 
    VQE: Variational Eigensolver; 
    varQITE: variational Imaginary Time evolition;
    Tortter Dynamics: Time evolution based on trotteried time propagation operator;
    TN: Tensor Networks;
    QTN: Quantum Tensor Networks inspired from classical TN; 
    varQTE: variational Quantum Real Time evolution;
    QNN: Quantum Neural Networks;
    QAOA: Quantum Approximate Optimization Algorithm; 
    HHL Algorith:  Quantum algorithm for linear systems of equations (by Aram Harrow, Avinatan Hassidim, and Seth Lloyd);
    QBM: Quantum Boltzman Machines;
    QCBM: Quantum Circuit Born Machine; 
    QGANs: Quantum Generative Adversarial Networks.
    See Appendix~\ref{app:algos_limits} for an overview of a selection of these methods.}
    \label{fig:qiskit-merged}
\end{figure}

This article is organized as follows. In Sec.~\ref{sec:ibm_roadmap}, we outline IBM's roadmap for future quantum devices and explain why digital quantum computers are suitable for addressing open challenges in \gls{hep}. Subsequently we describe the challenges in the field and goals that are one hopes to achieve utilizing quantum hardware in Sec.~\ref{sec:goals}. Section~\ref{sec:algs} contains a description of various algorithms that we consider as key candidates for achieving the goals outlined in the previous section. Finally, we conclude in Sec.~\ref{sec:conclusion_outlook}. In appendix~\ref{appendix_resources}, we provide a detailed estimation of the required resources for encoding lattice gauge theories in a digital, qubit-based quantum computer, while appendix~\ref{app:algos_limits} contains information on selected quantum and classical algorithms.

\section{IBM Roadmap on Quantum Computing\label{sec:ibm_roadmap}}

Bringing about useful quantum computing to the scientific world, and in particular, to the \gls{hep} community, is contingent on the development of quantum computing hardware and software that permits the execution of quantum algorithms at a scale that is capable of producing insights and results not accessible by classical computers. But more than only requiring a large-scale device, one requires that the components are sufficiently reliable and have coherence times as well as gate parameters of high quality~\cite{wack2021quality}. The IBM Quantum roadmap proposes a list of stepping stones that progressively improve on the necessary requirements. The first development roadmap was previewed in 2020~\cite{IBM_roadmap1} laying out a progression of the then available 27 qubits Falcon devices to the Condor chip with 1,121 qubits by the end of 2023. With the release of the 433 qubit Osprey chip a the end of 2022~\cite{OspreyIBMnews} the roadmap has been extended~\cite{IBM_roadmap2}. 
The new roadmap now lays out a path to the newly introduced Kookaburra chip with 4,105 qubits that utilizes interconnected chip designs with long-range couplers. Furthermore, the new roadmap added new chip architectures, such as the Heron chip with 133 qubits incorporating recent advances from gate and qubit research.
\begin{figure*}
    \centering
    \includegraphics[width=0.95\linewidth]{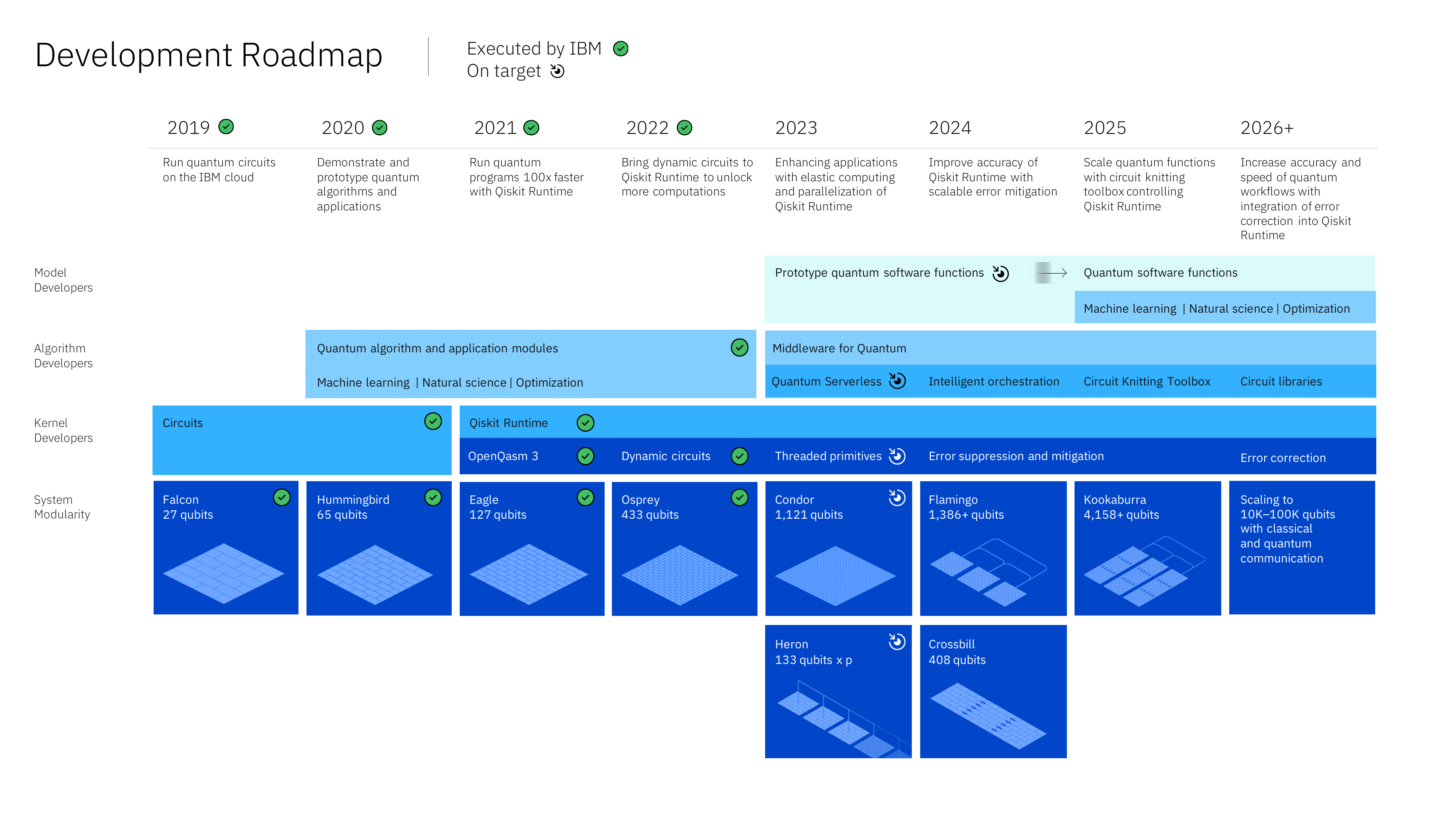}
    \caption{IBM's roadmap for upccoming quantum computers, updated 2022.}
    \label{fig:roadmap}
\end{figure*}

The greatest adversary to the realization of large-scale quantum computers is noise. The components of quantum computers are considerably more sensitive to imperfections and external interactions than their classical counterparts, leading them to decohere and turn into classical mixtures~\cite{unruh1995maintaining}. It is therefore almost universally accepted that complex and high-depth quantum algorithms such as Shor's factoring algorithm~\cite{shor1994algorithms}, quantum amplitude amplification~\cite{brassard1997exact,grover1998quantum}, phase estimation~\cite{kitaev1995quantum} or the long-time simulation of quantum dynamics will require quantum error correction. The design plans for the progressively larger \gls{qc} layouts are therefore aimed at providing a path to the long-term goal of realizing a fault-tolerant quantum computer. However, current error-correcting codes, which could be used to realize fault-tolerant quantum computing at a non-trivial scale, require system sizes that exceed the available hardware by several orders of magnitude~\cite{gidney2021factor,lee2021even}. Building a fault-tolerant computer, therefore, requires not only higher quality and larger scale devices but also research in error correcting codes. Recent advances in the theory of error correction~\cite{breuckmann2021quantum} provide us with reason to be optimistic about future progress. However, if we only wait for the realization of a fault-tolerant quantum computer to run algorithms and do not actively explore the potential of near-term devices, we will forgo a promising opportunity to obtain a computational advantage in the near future.

We are observing remarkable progress in quantum hardware. As the roadmap and the already completed milestones indicate, we are both building larger devices and can manufacture components with an order of magnitude improvement in two-qubit gate fidelities~\cite{Stehlik2021}. A quantum processing unit at the scale of the 65-qubit Hummingbird chip could implement circuits with a few thousand gates to a reasonable degree of accuracy without resorting to error correction when two-qubit gate fidelities of $99.99\%$ become available. Circuits of such a size can arguably no longer be simulated by exact methods on a classical computer. This suggests an alternative path of utilizing current and impending quantum devices~\cite{path_error_mit}. Here, one restricts to computations with only shallow-depth quantum circuits, where the size of the circuit is determined by hardware parameters such as coherence times and gate fidelities. As these parameters improve, the circuit sizes that become accessible increase, ultimately leading to circuits that provide a computational advantage over classical approaches. This path lays out a gradual progression to obtaining quantum advantage one hardware improvement at a time, ultimately driving the hardware evolution to progressively better and larger devices until error correction methods can be applied to provide us with access to circuits no longer limited by the device noise.

Early experiments~\cite{kandala2017hardware} demonstrated that despite the restriction to shallow-depth circuits, noise and decoherence lead to a bias in the estimates of expectation values. For this approach to provide an advantage over classical approximation methods this bias has to be mitigated. These observations have motivated the development of error mitigation tools such as \gls{zne}~\cite{Li2017Efficient,Temme2017Error} and \gls{pec}~\cite{Temme2017Error}. The goal of these methods is to reduce, or even fully remove, the noise-induced bias from expectation values measured in shallow-depth circuits. This is achieved by slightly modifying the circuits in different ways and combining measurement outcomes in post-processing to produce noise-free estimates. The protocols introduce an additional computational and sampling overhead that will ultimately grow exponentially in the noise strength, illustrating that these protocols do not extend the circuit depth beyond the device specific parameters, but only ensure that accurate values are produced within the allotted circuit size. The \gls{zne} method was experimentally implemented for the first time on small-scale chips~\cite{Kandala2019Error}. There it was shown that the effect of noise in earlier experiments~\cite{kandala2017hardware} could be removed. Recently it was demonstrated~\cite{Kim2022Scalable} that this method could be scaled to larger circuit sizes on improved quantum hardware, such as the recent version of the 27 qubit Falcon processor, by combining the method with error suppression techniques including dynamical decoupling~\cite{viola1998dynamical,viola1999dynamical} and Pauli - twirling~\cite{Bennett1996Purification,KnillFault2004,Kern2005Quantum}. Advances in learning and modelling correlated noise on quantum processors have enabled the implementation of \gls{pec}~\cite{Berg2022Probabilistic} to fully remove the noise bias for even the highest weight observables on larger devices.

To enable the scientific community to utilize these advances, IBM Quantum has announced a challenge to both internal developers as well as the community: the $100\otimes 100$ Challenge~\cite{IBM_100by100}. In 2024, IBM Quantum is planning to offer a quantum computing chip capable of calculating unbiased observables of circuits with 100 qubits and 100 depth of gate operations in a reasonable runtime, i.e. within a day. This new tool is to challenge the community towards proposing quantum algorithms that utilize this hardware to solve interesting problems, which are notoriously hard for classical computers. 

The \gls{hep} community plays a pivotal role here since the field is one of the driving sources for challenging computational problems inherent to quantum mechanics. This community is ideally equipped to propose problem relevant heuristics~\cite{RevModPhys.94.015004,Preskill2018} that stand to benefit from early demonstrations on quantum hardware 


\section{Challenges and Goals\label{sec:goals}}

\subsection{Selected Applications in the Theory Domain}
\label{subsect_Theory}

In this section, we will introduce a series of interesting theoretical challenges in different theoretical domains, including many-particle physics, different flavours of lattice gauge theories, and neutrino physics. The applications are dealing with relatively low dimensional systems, which however preserve some of the key aspects and criticality, which characterize the systems at the full scale. 
Clearly, the list of identified topics cannot be exhaustive. The choice is mainly motivated by the research interests of the co-authors of this paper. 
However, we hope that the solutions proposed for this selection of problems, and the corresponding algorithms, 
can be of inspiration in other domains not contemplated here.

Since most applications will deal with the dynamical aspects of the different model Hamiltonians, we will start this section with an introduction on methods for real-time simulations. 

\subsubsection{Simulations of Real-Time Phenomena}

Experimental results from high-energy physics labs, such as the Large Hadron Collider, come in the form of data on collision products. It is through scattering processes that we experimentally acquire a deep understanding of the fundamental physics, typically by reconstructing which composite quasiparticles are assembled during intermediate stages of the scattering event, and comparing their properties to the theoretical predictions from the standard model (and beyond).

It is clear, however, that this type of prediction presents several limitations. First of all, it is indirect, in the sense that the observed composite quasiparticle properties are compared, and not the scattering event distribution per se. Moreover, the analytic calculations of such quasiparticles are limited to those accessible via perturbation theory, in the form of Feynman diagrams, and thus no accurate predictions are expected for the quantum chromodynamics sector, which is far from perturbative. A substantial obstacle towards accurate model predictions of scattering phenomena is that Monte Carlo methods which excel at capturing equilibrium properties, are hindered when tackling out-of-equilibrium real-time dynamics, again, due to the sign problem and complex actions to numerically integrate.

From this perspective, gaining access to direct data of non-perturbative many-body real-time simulations of gauge theories would enable a complete paradigm shift. The simulation could immediately provide the statistics of products so that we could immediately compare them with the observed statistics of collected events from high-energy labs. Lattice gauge theories in the Hamiltonian formulation are perfectly suited for this task: while space dimensions are discretized (typically into a cubic lattice), time is kept as a continuous variable, and thus the many-body real-time evolution operator is formally well-defined for any arbitrary time interval. In this framework, the continuum limit can be safely approached without worrying about ultraviolet divergences.
Numerically computing such an evolution operator, and its action onto an arbitrary input state (for instance converging quasiparticle wavepackets) is however an exponentially hard problem in the lattice system size and requires the aid of quantum simulators or quantum-inspired numerical algorithms to be carried out in good approximation.

Both analog and digital quantum simulator strategies can be used to push towards this goal. In either case, the real-time evolution operator is applied to a set of qubits (or, more generally, \textit{qudits}) which encode the many-body quantum fields state. An analog quantum simulator approximates the target model Hamiltonian by implementing an instantaneous controllable Hamiltonian that is equivalent to the target at a chosen energy scale, and then lets the system evolve with time-independent controls~\cite{Banerjee2013}. This approach is inherently scalable, but it is limited by what types of interactions can be engineered. Conversely, digital quantum simulators aim at decomposing the action of the time evolution into a circuit of programmable quantum operations (for instance, gates)~\cite{Martinez2016}. This approach is more general, especially if the quantum resources form a universal set of gates, but it can be demanding in terms of scalability and coherence. Indeed, the number of qubits and the circuit depth required to perform such simulations are largely beyond the capabilities of current near term, noisy quantum 
devices~\cite{Lloyd96,Jordan2012}.

Alongside methods based on quantum hardware, we highlight the potential of Tensor Networks as a numerical strategy working on the same lattice Hamiltonian framework (discrete space, continuous time) as quantum simulators~\cite{Schollwock2011,Montangero2018a, Silvi2019a,Cirac2020}. \gls{tn} excel in describing lattice quantum states at equilibrium, even in multiple spatial dimensions, and even at finite densities~\cite{Dalmonte2016,Banuls2020TNreview,Felser2019,Magnifico2020}. Moreover, they can accurately capture out-of-equilibrium dynamics as long as the entanglement production is low (i.e.~the area laws of entanglement are not violated). While seemingly a strict requirement, it is actually a ubiquitous occurrence, from many-body localization, to slow quenches across phase transitions (Kibble-Zurek mechanism), to short-timescale transient phenomena under Lieb-Robinson bounds. Thus there are many physical systems whose dynamics are accurately captured by \gls{tn} (especially in one dimension).
Indeed, the first proof of principle demonstration of a scattering event in a lattice gauge theory in one-dimension was shown in~\cite{Pichler2016} where two-wave-packets collisions and subsequent time evolution of the created entanglement was studied. A more refined study of the process was presented in~\cite{Rigobello2021}. 

When specifically addressing scattering problems, with either classical or quantum simulations, there is an additional conceptual complexity which gets added to the already-serious problem of executing the many-body dynamics: namely, preparing the input state. Initial quantum states in particle colliders experiments typically involve localized wave packets of composite quasiparticles, for example hadrons. Written in the elementary quantum fields, these wave packets have a well defined center-of-mass momentum and overall number density (usually one quasiparticle), but their internal wave function can be very complex. Clearly, the scattering simulation must include strategies to build these states (and control their momentum) by carefully manipulating the elementary quantum fields encoded as qudits, starting from the (entangled) dressed vacuum.
Proposals to achieve such input-state preparation have been put forward for \gls{qtn}~\cite{VanDamme21,dborin2022} but the optimal general strategy is still unclear, and requires further investigation. Notice that this problem will remain when it becomes possible to study scattering processes in future quantum processors. Thus, any partial or final solution developed for tensor network will be highly valuable also for future quantum computations and the simulation of scattering processes. 
Let us mention in passing that other real-time phenomena, such as quenching, see e.g.~\cite{Banuls:2022iwk,Banuls2019b}, have also been studied with \gls{qtn} techniques. 

\subsubsection{(2+1)D QED}
\label{subsubsect_2+1QED}
As mentioned in the introduction, \gls{2p1D} \gls{qed} is one of the simplest quantum field theories that nevertheless retain interesting physics: for example it shares with \gls{qcd} important properties such as asymptotic freedom and confinement, and it is an excellent starting point for future analysis of more intricate theories.
We therefore propose \gls{2p1D} \gls{qed} as a very suitable benchmark and testbed model to explore the potential of quantum computing and, in particular, to compare it to \gls{tn} calculations.

The most used classical method to study lattice gauge theories numerically nowadays is the \gls{mcmc} approach, see the recent \textit{FLAG review}~\cite{FlavourLatticeAveragingGroupFLAG:2021npn}. While \gls{mcmc} can reach lattice sizes of order of $100^3\times 200$, which are currently unthinkable for \gls{qc} and \gls{tn} techniques, the Hamiltonian formulation used for the latter methods has several advantages.
For example, \gls{mcmc} suffers from very large autocorrelation times towards the continuum limit~\cite{Schaefer:2010hu}. In the regime of small to very small lattice spacing, we can take advantage of quantum computing or tensor network approaches that do not have this drawback. 
Furthermore, the Euclidean path integral used by \gls{mcmc} is afflicted by the infamous \textit{sign problem}~\cite{Troyer2005} which makes the study of quantum field theories at non-zero fermion densities impossible. More specifically for lattice \gls{qcd}, this prevents the exploration and characterization of regions of the phase diagram at non-zero baryon density, which are relevant to understand the early universe, neutron stars, or the transition to a quark-gluon plasma.  
Another important aspect is the limitation for classical \gls{mcmc} techniques in the presence of a topological term which, in stark contrast, can be treated  straightforwardly in the Hamiltonian formulation, i.e. with \gls{qc} or \gls{tn}. Finally, a Hamiltonian approach will enable the study of real-time phenomena such as scattering processes, thermalization or the dynamics of physical systems after quenching, see the discussion in Sec.~\ref{Introduction} and below.

Although we are fully aware of the advancements of \gls{tn}~\cite{Banuls2019SimulatingLG}, in the spirit of this paper, we will focus on the quantum computing approach to study quantum field theories and, in particular, on the example of \gls{2p1D} \gls{qed}.

 Another pillar of quantum information science and technology is analog quantum simulators~\cite{Hauke2013,Georgescu2014,Acin2018} which allow direct experimental access to various quantum many-body phenomena. Given recent advancements in quantum-simulator technology such as single-atom resolution through gas microscopes~\cite{Bakr2009,Bakr2010,Gross2021} and overall high levels of precision and control~\cite{Bloch2008}, quantum simulators have become an attractive venue on which to probe high-energy phenomena~\cite{Pasquans_review,Dalmonte2016,Zohar2015a,aidelsburger2021cold,zohar2021quantum}, affording the precious advantage of accessible temporal snapshots at any stage of the system dynamics. The \textit{modus operandi} of quantum simulators is to map a \textit{target model} described by a Hamiltonian $\hat{H}_0$ onto another quantum model amenable for realization in an experimental platform. This mapping is almost never exact but will lead to an effective model where $\hat{H}_0$ arises up to leading order in perturbation theory, along with (undesired) subleading terms $\lambda\hat{H}_1$, with strength $\lambda<1$. In the context of gauge theories, the model $\hat{H}_0$ hosts a gauge symmetry generated by local operators $\hat{G}_j$, while $\hat{H}_1$ explicitly breaks it.

Initially, quantum simulators of gauge theories were restricted to cold-atom realizations of building blocks for both $\mathbb{Z}_2$~\cite{Schweizer2019} and $\mathrm{U}(1)$ gauge groups~\cite{Mil2020}. The experiment of Ref.~\cite{Schweizer2019} employed two species of bosonic cold atoms in a double-well potential. Periodic driving resonant at the on-site interaction strength and with the appropriate fine-tuning of the modulation parameters resulted in an effective Floquet Hamiltonian with the desired $\mathbb{Z}_2$ gauge symmetry. On the other hand, the experiment of Ref.~\cite{Mil2020} employed inter-species spin-changing collisions to model the gauge-invariant coupling between matter and gauge fields. Although groundbreaking in their own right, these experiments were restricted to building blocks and suffered from uncontrolled subleading gauge-noninvariant processes that limited useful coherent times~\cite{Halimeh2020}.

To probe gauge-theory physics relevant to high-energy phenomena, it became essential to devise experimentally feasible methods that could enable large-scale implementations on quantum simulators. This was made possible through the introduction of \textit{linear gauge protection}. It could be shown that gauge violations were suppressed controllably up to all experimentally relevant timescales~\cite{Halimeh2021}. Such a term naturally arises in mappings of spin-$1/2$ representations of \gls{1p1D} lattice \gls{qed} on a tilted Bose--Hubbard superlattice, which has recently enabled the realization of a large-scale $\mathrm{U}(1)$ quantum link model on a quantum simulator composed of $71$ superlattice sites~\cite{Yang2020}. Stabilized gauge invariance was certified by adiabatically sweeping through Coleman's phase transition and observing a gauge violation of less than $10\%$ throughout the entire dynamics. This setup was then employed to study thermalization in the $\mathrm{U}(1)$ quantum link model~\cite{Zhou2022,Wang2022}, and further extended to probe rich quantum many-body scarring regimes in this gauge theory~\cite{Su2023}. Extensions of this large-scale platform with linear gauge protection have been proposed for higher spatial dimensions~\cite{Osborne2022} and for larger spin representations of the gauge field~\cite{osborne2023spins}.

In what follows and to be concrete, we consider the formulation of \gls{qed} on a two-dimensional space lattice with lattice spacing $a$.
Since the Hamiltonian formalism is to be considered for its eventual application on quantum devices, an encoding needs to be applied to represent the fermionic and gauge degrees of freedom, which cannot be fully eliminated in \gls{2p1D}.  To deal with the fermionic \textit{doubling problem}~\cite{NIELSEN1981219,PhysRevD.16.3031,rothe2012lattice}, i.e. the existence in $d$-dimensions of $2^d$ flavors (or tastes) for each physical particle, many different discretizations have been considered. One of the most used is the \gls{ks} formulation~\cite{Kogut1975}, which separates fermionic and antifermionic degrees of freedom and assigns them to alternate sites of the lattice. Therefore the fermions and antifermions are associated with a single component field operator $\hat{\phi}_{\vec{n}}$, with $\vec{n}=(n_x,n_y)$ as the coordinates of the lattice sites. The parity of the coordinate $n_x + n_y$ determines the type of matter associated to the site (i.e., with particles (antiparticles) placed on even (odd) sites).
The links of the lattice are identified by a site $\vec{n}$ and a direction $\mu = x,y$ emanating from that site. After introducing a proper discretisation of the U(1) group, such as $\mathbb{Z}_{2L+1}$ ($L \in \mathbb{N}$), the electric field operators for each link $\oper{E}_{\vec{n}, \mu}$ take integer eigenvalues $\oper{E}_{\vec{n}, {\mu}}\ket{e_{\vec{n}}}=e_{\vec{n}}\ket{e_{\vec{n}}}, e_{\vec{n}} \in \mathbb{Z}$. It is then necessary to truncate this number of eigenvalues to $(2l+1)$ ($l \in \mathbb{N}$ and $l\le L$), to represent the gauge fields on the (finite-size) quantum circuit. On the links, we also define the link operators
\begin{eqnarray}
\oper{U} = \mathrm{e}^{iag\oper{A}_{\mu}(\vec{n})},
\end{eqnarray}
where $\oper{A}_{\mu}(\vec{n})$ is the vector field and $g$ is the coupling constant. 
These operators obey the following commutator
\begin{equation}
[\oper{E}_{\vec{n}, \mu},\oper{U}_{\vec{n}^\prime, \nu}] = - \delta_{\vec{n},\vec{n}^\prime} \delta_{\mu,\nu}\oper{U}_{\vec{n}, \mu},
\end{equation}
and therefore act as a lowering operator on electric field eigenstates, namely $\oper{U}_{\vec{n}, {\mu}}\ket{e_{\vec{n}}}=\ket{e_{\vec{n}}-1}$. Physically $\oper{U}_{\vec{n},\mu}$ measures the phase proportional to the coupling acquired by a unit charge moved along a link. 

Setting the lattice spacing $a=1$, the Hamiltonian can thus be written as~\cite{PhysRevD.16.3031}:
\begin{equation}\label{eq:fullH}
\oper{H}_{tot} = \oper{H}_E + \oper{H}_B + \oper{H}_m + \oper{H}_{kin}.
\end{equation}

The first term is related to the electric interaction,
\begin{eqnarray}
	\oper{H}_E = \frac{g^{2}}{2} \sum_{\vec{n}}\left(\oper{E}^{2}_{\vec{n}, x} 
	+ \oper{E}^{2}_{\vec{n}, y}\right).
\end{eqnarray}

The second term in $\oper{H}_{tot}$ defines the magnetic interaction,
\begin{eqnarray}
\oper{H}_B = -\frac{1}{2g^{2}} \sum_{\vec{n}} \left(\oper{P}_{\vec{n}} + \oper{P}_{\vec{n}}^{\dag}
\right),
\end{eqnarray}
where $\oper{P}_{\vec{n}} =  \oper{U}_{\vec{n},x}\oper{U}_{\vec{n}+x,y}\oper{U}^{\dag}_{\vec{n}+y,x}\oper{U}^{\dag}_{\vec{n},y}$ is called plaquette operator.

The last two terms describe the fermionic part, i.e. the mass term 
\begin{eqnarray}
\oper{H}_{m} = m \sum_{\vec{n}} (-1)^{n_x+n_y} \oper{\phi}^\dag_{\vec{n}} \oper{\phi}_{\vec{n}},
\end{eqnarray}
with $m$ the fermion mass, and the kinetic term, corresponding to the creation or annihilation of a fermion-antifermion pair on neighbouring lattice sites,
\begin{eqnarray}
\oper{H}_{kin} = \sum_{\vec{n}} 
\frac{(-1)^{n_{xy}}}{2} (\oper{\phi}_{\vec{n}}^{\dag} \oper{U}_{\vec{n},
 x} \oper{\phi}_{\vec{n}+ x} + {H.c.}),
\end{eqnarray}
where  for the links in the $x$-direction $n_{xy}=1$ and for those in the $y$-direction $n_{xy}=(-1)^{n_{x}}$.

An alternative to the \gls{ks} formulation is the \textit{Wilson approach}~\cite{PhysRevD.10.2445,angelides2023computing}. It introduces a second-order derivative term in the Hamiltonian that vanishes linearly with the lattice spacing in the continuum limit.
The main advantage of this approach is that the number of qubits needed to represent the gauge fields is lower than the one utilised in the \gls{ks} approach, and therefore has a lower resource requirement~\cite{Mathis2020}.

One of the challenges of simulating the gauge theory with quantum computers is to find a resource efficient way to map all its degrees of freedom onto a quantum computer. This holds, in particular, for the bosonic gauge degrees of freedom. Here several Ans\"atze exist in the 
literature~\cite{Clemente2022a,Bauer:2021gek,Haase2021resourceefficient,Kane:2022ejm} and it is important to test these approaches against each other, evaluate their advantages and shortcomings, and identify the most resource efficient discretization and truncation scheme for their implementation on a quantum computer.  

Once we have developed the most suitable encoding, we need to choose the most appropriate simulation technique depending on our goal. For example, in order to compute the ground state energy (the low-lying spectrum) of our Hamiltonian we can apply \gls{vqe}~\cite{Peruzzo2014} (\gls{vqd}~\cite{Higgott2019} or \gls{ssvqe}~\cite{PhysRevResearch.1.033062}).
Other approaches could be imaginary time evolution~\cite{McArdle_2019} or creating a suitable operator basis~\cite{PhysRevResearch.2.043140}. 

\subsubsection{(2+1)D SU(2)}

With the long-term goal of quantum chromodynamics in mind, it is important to consider non-Abelian gauge theories. A Yang-Mills theory with SU(2) gauge symmetry group is a natural first step. The standard Kogut-Susskind Hamiltonian formulation of lattice gauge theories is defined as 

\begin{equation}
    \begin{aligned}
        \hat{H} &= \frac{1}{2a} \sum_{\vec{n}} \sum_{\alpha,\beta} \left(i \hat{\psi}_\alpha (\vec{n})^{\dag}\hat{U}_{\alpha\beta}(\vec{n},i)\hat{\psi}_\beta(\vec{n}+\hat{i})\right.\\
         &\phantom{ \frac{1}{2a} \sum_{\vec{n}} \sum_{\alpha,\beta}}\left.+ (-1)^{\vec{n}} \hat{\psi}_\alpha (\vec{n})^{\dag}\hat{U}_{\alpha\beta}(\vec{n},j)\hat{\psi}_\beta(\vec{n}+\hat{j})
        + \text{H.c.} \right) \\
        & + m\sum_{\vec{n}} \sum_{\alpha} (-1)^{\vec{n}} \hat{\psi}^{\dag}_{\alpha}(\vec{n})\hat{\psi}_{\alpha}(\vec{n})\\
        & + \frac{g^2}{2a^{d-2}} \sum_{\vec{n},l}\sum_{b} [\hat{E}^b(\vec{n},l)]^2 \\
        & -\frac{1}{2a^{4-d}g^2}\sum_{\vec{n}} \sum_{\alpha, \beta, \gamma, \delta}\left(\hat{U}_{\alpha \beta}(\vec{n},i)\hat{U}_{\beta \gamma}(\vec{n}+\hat{i},j)\right. \\
        & \phantom{-\frac{1}{2a^{4-d}g^2}\sum_{\vec{n}} \sum_{\alpha, \beta, \gamma, \delta}}\left.\times\hat{U}_{\delta \gamma}^{\dag}(\vec{n}+\hat{j},i)\hat{U}_{\alpha \delta}^{\dag}(\vec{n},j) + \text{H.c.}\right).
    \end{aligned}
    \label{KS_Ham}
\end{equation}
where $\vec n$ is a lattice site and $l$ is a direction on the spatial lattice. Greek indices such as $\alpha=1,2$ (and $\beta$, $\gamma$, $\delta$) are indices in the fundamental representation of the SU(2) group, whereas $b=1,2,3$ is an adjoint index. Physically, $\hat E^b$ is the chromoelectric field and, as discussed for \gls{qed}, the chromomagnetic field arises from the plaquette term that appears last in Eq.\eqref{KS_Ham}. The physical parameters in this Hamiltonian are the fermion mass $m$ and the gauge coupling $g$.

The problem of simulating lattice gauge theories on a universal quantum computer using qubits as the basic degrees of freedom was defined in general terms in~\cite{Byrnes2006}. It was shown in~\cite{kan2021lattice} that lattice gauge theories in any spatial dimension can be simulated on quantum hardware using a polynomial number of gates in the number of lattice sites, bosonic gauge field truncation, and simulation time.

Other approaches use quantum simulators to emulate the physics of non-abelian gauge theories. In these implementations, gauge invariance is a direct consequence of some underline symmetry of the quantum simulator. For instance, angular momentum conservation is used to realise the SU(2) Yang-Mills model~\cite{Zohar2013a} and nuclear spin conservation in alkaline-earth atoms is used to mimic SU(N) models within the quantum link formulation~\cite{Banerjee2013}. In this respect, the quantum link formulation appears as a natural formulation for the quantum simulation of the non-abelian model which was proposed within a Rydberg-based architecture~\cite{Tagliacozzo2013a} and within superconducting circuits~\cite{mezzacapo2015non}.

The first quantum simulation of a SU(2) lattice gauge theory on IBM superconducting hardware was done in~\cite{Klco2019}. Subsequently, exploratory computations were conducted for one-dimensional SU(2) on an IBM superconducting platform~\cite{Atas2021}. This implementation combined the fact that a one-dimensional theory with open boundary conditions allows one to rewrite all gauge field degrees of freedom as long-range interactions among fermions with \gls{vqe} to study both meson and baryon states. There have further been one-dimensional SU(3) quantum simulations~\cite{farrell2022preparations,atas2022real,farrell2022preparationsb}, and error mitigation methods have been applied to study the time evolution of non-abelian models~\cite{rahman2022self}. 

In contrast to the one-dimensional case, studies of two-dimensional SU(2) gauge theory require both fermion and gauge field degrees of freedom. Several formulations have been proposed~\cite{raychowdhury2020loop,davoudi2021search}, and practical studies of each will provide valuable information for understanding their advantages and disadvantages.

The choice of basis for the local degrees of freedom in the implementation of a non-abelian gauge model is important~\cite{davoudi2022general,tong2022provably}. Usually, one needs to study the effect of discretisation or truncation on the physical results of the models~\cite{hartung2022digitising,jakobs2023canonical}. One option to discretise non-abelian theories is to use finite-dimensional subgroups~\cite{gustafson2022primitive,zohar2017digital,bender2018digital}, which can be efficiently implemented within a Rydberg base architecture~\cite{gonzalez2022hardware,gonzalez2023fermionic,zache2023quantum,zache2023fermion}. Early computations of SU(2) gauge fields on quantum hardware have used lattices with up to 6 plaquettes in total~\cite{Klco2019,rahman20212}. An initial study was also carried out for SU(3) in~\cite{Ciavarella2021}.  

Upcoming computations can build upon the lessons learned from these first steps, and grow in scale and scope alongside the continuing progress in quantum hardware deployment in the noise intermediate scale quantum era.

\subsubsection{Quantum Link Models and D-Theory}

D-theory is an alternative formulation of lattice field theory in which continuous, classical fields are replaced by {\em discrete}, quantum degrees of freedom, which undergo {\em dimensional reduction} from an extra dimension of short extent \cite{brower2004d}. In the D-theory approach, lattice gauge theories are realized via quantum link models
\cite{horn1981finite,orland1990lattice,chandrasekharan1997quantum,brower1999qcd,wiese2013ultracold,wiese2014towards}. Quantum links are generalized quantum spins endowed with an exact gauge symmetry, which is located on the links of a spatial lattice.

Quantum links reside in finite-dimensional irreducible representations of an embedding algebra. This is in contrast to the standard Wilson-type lattice gauge theory, which is based on an infinite-dimensional representation on each link. Quantum links with a $U(N)$ or $SU(N)$ gauge group reside in the embedding algebra $SU(2N)$. In particular, $U(1)$ quantum link models are formulated with ordinary $SU(2)$ quantum spins. $SO(N)$ and $Sp(N)$ quantum link models are realized with an $SO(2N)$ and $Sp(2N)$ embedding algebra, respectively. Since $SU(2) = Sp(1)$, an $SU(2)$ quantum link model can be realized with a simple $Sp(2) = SO(5)$ embedding algebra.
 
The simplest Abelian $U(1)$ quantum link model is realized with ordinary quantum spins $1/2$, which can be embodied by individual qubits. These dynamics have already been represented by quantum circuits in a resource-efficient manner \cite{wiese2022quantum}. The implementation of the $U(1)$ quantum link model on a triangular lattice is particularly simple, because it takes advantage of the heavy hexagonal lattice topology underlying the 127-qubit IBM Eagle chip \cite{banerjee2022nematic}. 

The simplest non-Abelian $SU(2)$ quantum link model uses the embedding algebra $Sp(2) = SO(5)$, which has a 4-dimensional fundamental representation that can be embodied by a pair of qubits residing on each lattice link. By an exact duality transformation, this $SU(2)$ quantum link model can be expressed in terms of $Z(2)$-valued height variables, which can even be embodied by individual qubits \cite{banerjee2018s}. This model is also interesting from a condensed matter perspective, because it is closely related to the quantum dimer model on the Kagom\'e lattice which has a rich, non-trivial phase structure. It would be very interesting to construct a quantum circuit, similar to the one for the $U(1)$ quantum link model on the triangular lattice, in order to perform quantum computations of the real-time dynamics of $SU(2)$ gauge theories.

\subsubsection{(1+1)D $\CP(N-1)$ Models from $(2+1)$D $SU(N)$ Quantum Spin Ladders}

\gls{1p1D} $\CP(N-1)$ quantum field theories are toy models that share many important features with \gls{3p1D} \gls{qcd}: they are asymptotically free,  have a non-perturbatively generated massgap, as well as $\theta$-vacua \cite{d19781n,eichenherr1978n}
In addition, they have non-trivial phase structure at non-zero chemical potential, including Bose-Einstein condensates with and without ferromagnetism \cite{evans20183}. 

The standard lattice formulation of $\CP(N-1)$ models at non-zero vacuum angle or at non-zero chemical potential suffers from similar sign and complex action problems as \gls{qcd} itself.
D-theory offers an alternative approach to standard lattice field theory, which uses {\em discrete} quantum (rather than continuous classical) degrees of freedom without compromising exact continuous symmetries including gauge symmetry. In asymptotically free theories (including \gls{1p1D} 
$\CP(N-1)$ models and \gls{3p1D} \gls{qcd}, the continuum limit is reached naturally (i.e.\ without any fine-tuning) via {\em dimensional reduction} from a higher-dimensional space-time, with a short extent of the extra dimension.
Interestingly, the finite-density and $\theta$-vacuum sign problems of the standard formulation have already been overcome by the alternative D-theory formulation, in which $\CP(N-1)$ models are regularized using $SU(N)$ quantum spin degrees of freedom \cite{beard2005study}
This formulation is also amenable to analog quantum simulations with ultra-cold alkaline-earth atoms in  optical lattices, which holds the promise to facilitate real-time simulations of their dynamics \cite{laflamme2016cp}
$\CP(N-1)$ models in the D-theory formulation are ideally suited as a testing ground for quantum computation, because, on the one hand, at least in some cases, advanced classical computational techniques are available  for validation, and, on the other hand, similar methods can be developed for lattice gauge theories, ultimately aiming at \gls{qcd}, in particular in the quantum
link formulation. 
The strategy behind D-theory, namely to formulate quantum field theory directly in terms of quantum degrees of freedom, is ideally suited for both quantum simulation and quantum computation.

The standard formulation of $\CP(N-1)$ models uses classical, Hermitean, idempotent $N \times N$-matrix fields $P(x)$
\[P(x)^\dagger = P(x) \ , \quad P(x)^2 = P(x) \ , \quad \mbox{Tr} P(x) = 1 \ , \]
with the Euclidean action
\[S[P] = \int d^2 x  \frac{1}{g^2} \mbox{Tr}\left[\p_\mu P \p_\mu P\right] + i \theta Q[P] \ ,\]
and the integer-valued topological charge
\[Q[P] = \frac{1}{\pi i} \int d^2 x \, \epsilon_{\mu\nu} \mbox{Tr} \left[P \p_\mu P \p_\nu P\right] \in \Pi_2[\CP(N-1)] = \Z \ .\]
The model is invariant under a global $SU(N)$ symmetry, $P(x)' = \Omega P(x) \Omega^\dagger$, $\Omega \in SU(N)$. 

The alternative D-theory formulation replaces the classical field $P(x)$ by $SU(N)$ quantum spins $T^a_x$ ($a \in \{1,2,\dots,N^2-1\}$) that obey the commutation relation
\[[T_x^a,T_{x'}^b] = i \delta_{xx'} f_{abc} T_x^c \ ,\] and reside on a 2-d spatial square lattice (of spacing $a$) with a long  $x_1$-direction (of extent $L$ with periodic boundary conditions) and a short $x_2$-direction (of extent $L'$ with open boundary conditions).
The even-parity sites $x \in A$ (with even $x_1 + x_2$) carry the fundamental representation 
$\{N\}$, $T_x^a = \lambda^a/2$ (where the $\lambda^a$ are Gell-Mann matrices), while the odd-parity sites $y \in B$ carry the anti-fundamental representation $\{\overline{N}\}$, $\overline{T}^a_y = - {\lambda^a}^*/2$. 
An antiferromagnetic $SU(N)$ quantum spin ladder (with $J > 0$) is then described by the 
nearest-neighbor Hamiltonian
\[H = J \sum_{\langle xy \rangle} T^a_x \overline{T}^a_y \ ,\] which commutes with the total $SU(N)$ spin $T^a = \sum_{x \in A} T^a_x + \sum_{y \in B} \overline{T}^a_y$.
In the presence of chemical potentials $\mu_a$ at inverse temperature $\beta$
the grand canonical partition function then takes the form
\[Z = \mbox{Tr} \exp(- \beta (H - \mu_a T^a)) \ .\]

Remarkably, this antiferromagnetic quantum spin ladder is a proper regularization for the \gls{1p1D} $\CP(N-1)$ quantum field theory. An even extent $L'/a$ of the short dimension corresponds to vacuum angle $\theta = 0$, while an odd extent implies $\theta = \pi$. For $L = L' = \beta = \infty$, the quantum antiferromagnet breaks the global $SU(N)$ symmetry down to $U(N-1)$ (at least for $N \leq 4$). 
This gives rise to dynamically generated, effective Goldstone boson fields $P(x)$ that reside in the coset space $SU(N)/U(N-1) = \CP(N-1)$.
Once $L'$ is made finite, the Mermin-Wagner theorem implies that $SU(N)$ can no longer break spontaneously. As a result, the previously massless Goldstone bosons pick up an exponentially small mass proportional to $\exp\left(- 4 \pi L' \rho_s/c N\right)$, where $\rho_s$ is the spin stiffness and $c$ is the spinwave velocity. 
For moderately large $L'/a \gtrsim 4$, the corresponding correlation length $\exp\left(4 \pi L' \rho_s/c N\right) \gg L'$ exceeds the extent of the short dimension and the system dimensionally reduces to the \gls{1p1D} $\CP(N-1)$ model. 
These dynamics, which may seem complicated at first glance, have been verified in great detail in quantum Monte Carlo simulations using classical computers. Already at the level of classical computation, the use of discrete quantum, rather than continuous classical, fundamental degrees of freedom has led to numerous algorithmic advantages, which facilitated efficient numerical simulations of $\theta$-vacua and
dense matter systems \cite{beard2005study,evans20183}

Analog quantum simulators for the $SU(N)$ quantum antiferromagnet have already been designed, using ultra-cold alkaline-earth atoms in an optical lattice, and are ready to be realized in the laboratory already today
\cite{laflamme2016cp}. 
This holds the promise to address the real-time dynamics, which remains inaccessible to classical simulation techniques. This would be the first time that an asymptotically free quantum field theory is studied with quantum simulation.
The simple nature of the quantum spin degrees of freedom and the ultra-local form of their Hamiltonian strongly suggest to also explore $\CP(N-1)$ models using digital quantum computation. 
In particular, in D-theory the $\CP(1)$ model with a global $SU(2)$ symmetry is regularized with ordinary $SU(2)$ quantum spins which can be embodied directly by individual qubits. 
Similarly, the $SU(3)$ quantum spins in the D-theory formulation of the $\CP(2)$ model are nothing but qutrits. 
The corresponding Hamiltonian dynamics can be realized with sequences of single-qubit and two-qubit (or single-qutrit and two-qutrit) quantum gates.
It is possible --- and already quite interesting --- to work with quantum spin chains (i.e.\ with $L'/a = 1$) rather than with quantum spin ladders ($L'/a > 1$).
In particular, for $L'/a = 1$ the antiferromagnetic  $SU(2)$ quantum spin chain corresponds to the \gls{1p1D} Wess-Zumino-Novikov-Witten conformal quantum  field theory in the continuum limit.
The corresponding $SU(3)$ quantum spin system, although it is not in the continuum limit, describes a strongly coupled $\CP(2)$ model at a first-order phase transition with spontaneously broken 
charge conjugation symmetry. 
This would allow, for example, real-time studies of false vacuum decay.

\subsubsection{Collective Neutrino Oscillations}
 
Neutrinos play a central role in extreme astrophysical events like core-collapse supernovae and neutron star binary-mergers as they dominate the transport of energy, entropy and lepton number. Due to the fact that neutrinos have masses and that 
the mass basis, denoted by $\{ | \nu_i \rangle\}_{i=1,3}$, is different from the flavor basis, neutrinos will experience oscillations in the population of the different flavors  components $(\nu_e, \nu_\mu , \nu_\tau)$. 

Given the importance of charge-current reactions, a detailed understanding of flavor oscillations in these settings is critical to predict their dynamical evolution. Given the high density of neutrinos in these environments, flavor oscillations are strongly affected by two-body neutrino-neutrino interactions, which render the neutrino cloud a strongly coupled many-body system. Direct solution of the evolution equations for general initial conditions can be exponentially hard with classical simulations, and the conventional approach is to rely on mean-field approximations~\cite{Duan2006,Duan2010,Tamborra2021}, which, however, do not include direct scattering between neutrino. Efforts in going beyond mean-field with classical computers were recently reviewed in~\cite{Patwardhan2023}. 

The complexity of neutrino physics 
persists even with the simplifying assumption that only two flavors (the electron flavor $\nu_e$ and one heavy flavor $\nu_x$) participate in the oscillation. With this assumption, one can model each neutrino as a set of interacting two-level systems and obtain the following Hamiltonian~\cite{Pehlivan2011}
\begin{equation}
\label{eq:ham_neutrinos}
H=\sum_{i=1}^N \boldsymbol{b}_i \cdot \boldsymbol{\sigma}_i + \lambda_e \sum_{i=1}^N \sigma^z_i + \frac{\mu}{2N}\sum_{i<j}^N \left(1-\cos(\theta_{ij})\right) \boldsymbol{\sigma}_i \cdot \boldsymbol{\sigma}_j\;, 
\end{equation}
where $\boldsymbol{\sigma}_i=(\sigma_i^x,\sigma_i^y,\sigma_i^z)$ is a vector of Pauli matrices acting on the i-th neutrino.
The first term in Eq.~\eqref{eq:ham_neutrinos} describes vacuum oscillations around the mass basis with $\boldsymbol{b}_i = \frac{\delta m^2}{4 E_i}( \sin(2 \theta_{\nu}), 0, - \cos(2\theta_{\nu}))$,
where $\delta m^2 = m_2^2 - m_1^2 $ is the square mass difference between mass eigenstates, $\theta_\nu$ is the mixing angle and $E_i$ the energy of the $i$-th neutrino. The second term in Eq.~\eqref{eq:ham_neutrinos} is generated by charge-current scattering with a background of electrons with coupling constant $\lambda_e=\sqrt{2}G_Fn_e$ with $G_F$ the Fermi constant and $n_e$ the electron density. This is the term responsible for the \gls{msw} effect due to the interaction of electrons with neutrinos experienced by neutrinos travelling in dense matter. Finally, the third term in Eq.~\eqref{eq:ham_neutrinos} is the neutrino-neutrino interaction generated by neutral-current weak reactions. Its coupling constant $\mu=\sqrt{2}G_Fn_\nu$ is directly proportional to the local neutrino density $n_\nu$, while the angular factor inside the sum encodes the spatial geometry of the problem through its dependence on the relative angle of propagation $\cos(\theta_{ij})=\boldsymbol{p}_i\cdot\boldsymbol{p}_j/(\|\boldsymbol{p}_i\|\|\boldsymbol{p}_j\|)$ 
where $\boldsymbol{p}_i$ is the momentum of $i$-th neutrino. This term prevents collinear neutrinos from interacting.

The Hamiltonian in Eq.~\eqref{eq:ham_neutrinos} can be used to describe the flavor evolution of a homogeneous gas of neutrinos at fixed density. Most neutrinos are however leaving the explosion region of the emitter (neutron stars, ...) where they have been generated and thus experience different local conditions as they move out. This can be incorporated by allowing the coupling constants $\lambda_e$ and $\mu$ to change with the distance $r$ from their emission, or equivalently with the time $t$ since they left the neutrino sphere (neutrinos are considered as ultra-relativistic particles moving at approximately the speed of light). With this, we are left with describing the non-equilibrium evolution of a large number of fermions interacting through an eventually non-adiabatic two-body Hamiltonian. Several extensions are possible to account e.g. for the full three-flavor structure~\cite{Siwach2022} or the presence of inhomogeneities~\cite{Stirner_2018} but are likely beyond the scope of the $100 \otimes 100$ IBM challenge. 

Current efforts to study the full many-body flavor dynamics generated by the Hamiltonian in Eq.~\eqref{eq:ham_neutrinos} beyond the mean-field approximation have been carried out under a number of additional simplifying assumptions. A popular one is to consider an average interaction strength, effectively removing the angular dependence in the two-body interaction turning it into a term proportional to the square of the total angular moment. This has the effect that the system becomes integrable using the Bethe-Ansatz~\cite{Pehlivan2011} and classical simulations have been performed in the past exploiting directly this property up to $N=9$~\cite{Cervia2019}. Using more direct integration approaches allowed $N=16$ to be reached~\cite{Patwardhan2021} while using \gls{mps} together with the Time Dependent Variational Principle systems up to $N=20$ were studied while keeping a good convergence with the bond dimension~\cite{Cervia2022}. Note that the latter simulations employed around $10^5$ time steps for the entire calculation and this leaves a direct comparison possibly out of range of the $100 \otimes 100$ IBM challenge.

Another common assumption is to neglect the MSW term proportional to the electron density using the argument that this coupling constant greatly dominates in the interior regions where many-body effects are expected to be important. The MSW term can be eliminated using a rotating wave approximation which ultimately produces a lower effective mixing angle. In general this is not necessary as this one-body term can be trivially fast-forwarded and included correctly, and efficiently, in the simulation by resorting to interaction picture schemes like the one proposed in~\cite{Rajput2022hybridizedmethods}. In the absence of this term, the Hamiltonian enjoys a global $U(1)$ symmetry generated by rotations around the mass basis which can be used to reduce the implementation cost. This strategy was used in~\cite{Yeter2022}, together with the use of IBM Qiskit’s isometry function to implement evolution in each sub-block, to study flavor oscillations up to  $N=4$ neutrinos systems. The approach has the advantage that the circuit depth does not increase as a function of the time steps but for large system sizes it would require an exponentially large number of gates. Part of the difficulty in including the two-body interaction is its all-to-all nature which naively does not fit well on devices with reduced connectivity. The problem can however be circumvented using an appropriate SWAP network scheme producing a circuit with $N$ layers of $~N/2$ nearest neighbor two-qubit gates each. This approach was proposed in~\cite{Hal21} where a $N=4$ neutrino simulations with a single Trotter step was carried out on IBM devices and has been shown to be advantageous to allow for classical simulations using \gls{mps}~\cite{Roggero2021}. Platforms that allow for all-to-all connectivity,  like trapped-ions, allow more flexibility but require a similar number of two-qubit operations. Due to their current higher fidelity, simulations have been reported for up to $10$ time steps with $N=4$ and for one time step up to $N=12$ neutrinos~\cite{Amitrano2022,Illa2022b}. Approaches using quantum annealers have also been proposed and applied for systems up to $N=4$~\cite{Illa2022a}.

The simplified neutrino oscillation problem described by Eq.~\eqref{eq:ham_neutrinos} is encoded quite naturally on a digital quantum computers with one qubit per neutrino. Current attempts to describe neutrinos on these platforms are still restricted to small $N$ values with rather simple initial conditions, usually wave-functions describing non-correlated neutrinos. Besides the description of larger neutrino number, challenges for future applications include the extension to more realistic initial conditions like initially thermalized neutrinos, or the evolution of these correlated systems over longer time to extract for instance asymptotic entanglement between neutrinos or characterize the relaxation dynamics to thermal states~\cite{Martin2023}. Time-evolution requires efficient algorithms to simulate the dynamics (see section~\ref{subsubsec_algo_dyn}). A first order \gls{pf} step for $N$ neutrinos costs $3N(N-1)/2$ \gls{cnot} operations while a second order step will cost $3(N^2-3N/2+1)$ \gls{cnot} gates~\cite{Amitrano2022}. The depths are instead $3N$ and $6N-3$, respectively. The implementation can be performed in a more hardware efficient way employing cross-resonance gates instead at the price of increasing the decomposition error. 
Furthermore, a hardware friendly approach to multi-product formulas can further reduce circuit depth and increase simulation accuracy~\cite{Vazquez2022}.
An additional possibility worth pursuing in the short term is the use of approaches based on \gls{vte} which allow for a circuit depth independent on the evolution time.

\subsection{Selected Applications for Experiments}
\label{subsect_Experiments}

High-energy physics experiments are characterized by the need to process a large amount of complex, highly structured data. Historically, large collaborations have relied on massively parallel computing infrastructure and pioneered the field of distributed computing with the LHC Computing Grid.
The need to search for processes with small production cross section together with next generation detectors,  generate a sheer size of the data sets to analyze, that require a new computing model, more efficient algorithms -- including data-driven techniques such as artificial intelligence--, and the integration of new hardware beyond the von Neumann architectures.
It is in this context that investigations about the introduction of quantum computing in \gls{hep} experiments is framed: the community is looking into accelerating or improving the different steps of data analysis and data processing chains.
Currently most of the work is focused on the development and optimisation of \gls{qml} algorithms implemented either as quantum neural networks (variational algorithms) or kernel methods~\cite {guan2021quantum, delgado2022quantum}. See Appendix~\ref{app:algos_limits} for a summary of these methods. The next section will give an overview of the range of algorithms under study as applied to HEP and their present limitations.

It is important to notice, however, that evaluating the performance of \gls{qml} algorithms on \gls{hep} data requires care: realistic applications have requirements that can not be easily accommodated on quantum devices, today.
The most critical issue is related to the size of data samples, together with their complexity.
Indeed, studies on the introduction of quantum algorithms (and \gls{qml} in particular) need to take into account both the total number of events that need analyzing (that can easily reach hundreds of thousands) and the large number of input features in each single event (typically in the order of tens or hundreds). The preferred approach today is hybrid:  a classical feature extraction and/or dimensionality reduction step is used to bring the classical input to a size that can be realistically embedded on 
noisy, near-term quantum hardware. 
Depending on the complexity of both the dataset and the task,  different methods are used, ranging from linear PCA, to non-linear  trainable embedding or compression methods (auto encoders or other AI-based techniques)~\cite{albertsson2019machine}. The advantage of the latter is clearly their versatility and the possibility to train them together with the quantum algorithms for the specific task at hand.

In particular, trainable techniques allow an end-to-end optimisation of the reduced data representation (often referred to as “latent representation”), their embedding in quantum states and the quantum algorithm itself. A binary classification problem, such as the separation of  signal versus uninteresting background, is a common example: simultaneously training an autoencoder for data compression together with the corresponding classifier, ensures that the resulting latent representation exhibits maximal separation between the two classes.
Multiple examples have already proven the advantage of this approach in both the classical and quantum domain~\cite{collins2022machinelearning, Farina_2020, Cheng_2023, QADCERN, wozniak2023quantum,belis2021higgs}
In addition, a critical part of the quantum algorithm design and optimisation process is  aimed at reducing the number of input features needed by the quantum algorithm in order to perform its task, together with the definition of a minimal training set, that still ensures convergence and generalization capabilities. 

Finally, the compressed classical data is embedded, or loaded, onto quantum states for processing by the \gls{qml} algorithm. This step is commonly referred to as the state preparation step.  Different  techniques have been studied~\cite{lloyd2020quantum} that compromise between an optimal use of qubit states, exploiting in full the potential exponential advantage, and the need to efficiently map state preparation circuits on 
noise devices. In general, the choice of the data embedding strategy has an effect both on the performance of the overall algorithm and on its interpretation (as, for example, in the kernel formalism) as mentioned in Sec.~\ref{sec:algs}.

Taken all together, these  steps  have made possible the design and implementation of quantum algorithms for most of the tasks in the typical data processing chain, albeit at a reduced scale. Access to the $100 \otimes 100$ quantum hardware, combined to data reduction techniques is likely to bring current prototypes to a much more realistic size.

\subsubsection{Rare Signal Extraction}
Extracting rare signals from background events is an essential part of data analysis in the search for new phenomena in \gls{hep} experiments. In this section we will cover algorithms, methods and limitations of this area of research, giving some references which, for sure, do not represent a complete picture of the state of art.

Posed as a classification task, rare signal extraction faces an imbalance problem in the number of samples belonging to the signal class versus the number of samples from the background class.
Entry level cases are the ones where a single feature is powerful enough to discriminate the process of interest while more complicated cases rely on multi-variate analysis of many features to get to a reasonable level of discrimination power.

In the machine learning community, techniques for learning from imbalanced data are well established, and for the \gls{hep} case, analysis methods developed in~\cite{Britsch:2008mxb, extr_imb} have been effectively implemented. An alternative approach to classification with imbalance techniques is anomaly detection~\cite{Edelen:2021vcs,anoma2}.
In the following we touch upon some modern class imbalance techniques adopted in the community, focusing on novel loss functions and data re-sampling techniques.
However, the main goal here is not merely the classification task but also the generation of predictions with their corresponding uncertainties. In particle physics, as in other scientific domains, if uncertainties are not presented the picture is almost incomplete.

Using the accuracy of a classifier as a metric for rare events can be misleading as it says nothing about the signal, in terms of distribution and feature importance. The ROC curve is a good general purpose metric, providing information about the true and false positive rates across a range of thresholds, and the area under the ROC curve (AUC) is a good general purpose single number metric.
Nevertheless, when dealing with imbalanced data the precision-recall curve is the preferred metric, where the recall represents a measure of how many true signal events have actually been identified as signal and precision quantifies how likely an event is to truly be signal and depends on how rare the signal is. Different strategies can be used like under-sampling the majority class or oversampling the minority class, where the former is preferred because of the potential overfitting resulting from oversampling~\cite{Nguyen2009BorderlineOF}. Moreover, the standard algorithm can be modified by playing with the hyperparameters of the loss and by adding an additional penalty for misclassification. For instance, following~\cite{focalLF}, a modified version of the cross entropy loss function used for binary classification to differentiate between easy- and hard-to-classify samples is the focal loss function: 
\begin{equation}
FL = -(1-p_t)^\gamma \log(p_t) \, .
\end{equation}
Here $p_t$ is the model’s estimated probability that a given event belongs to the signal class and $\gamma$ is the modulating parameter. As $\gamma$ is increased the rate at which easy-to-classify samples are down weighted also increases.
As pointed out previously, not only should the classification be efficient but also the related prediction uncertainties.
Current approaches for this include dropout training in Deep Neural Networks as approximate Bayesian inference, variance estimation across an ensemble of trained deep neural networks, and Probabilistic Random Forest~\cite{modeluncertainty}.
For example, such techniques have been used in for the measurement of the longitudinal polarization fraction in same-sign $WW$ scattering~\cite{Ballestrero_2018} and for the decay of the Higgs boson to charm-quark pairs~\cite{Aaboud_2018}.

Same-sign $WW$ production at the LHC is the \gls{vbs} process with the largest ratio of electroweak-to-\gls{qcd} production. As such it provides a great opportunity to study whether the discovered Higgs boson leads to unitary longitudinal \gls{vbs}, and to search for physics \gls{bsm}.
Confirming or refuting the unitarity of \gls{vbs} requires not just a measurement of $pp\to jjW\pm W\pm$, but of the fraction of these events where both $Ws$ are longitudinally polarized (LL fraction). The fraction of longitudinally polarized events is predicted to be only a fraction $ 0.07$ to the total number of events in the \gls{sm} at large dijet invariant mass $(m_{jj})$~\cite{Ballestrero_2018} making this a challenging measurement.
Common techniques for this kind of use cases are Random Forest with imbalanced implementation, Gradient Boosted Decision Tree and Deep Learning models with standard or focal loss function. Overall, all of the machine learning models significantly outperform the kinematic variables approach~\cite{Grossi_2020}. 

The second application of class imbalance techniques is the measurement of Higgs boson decays to charm-quark pairs. Searches for the decay of the Higgs boson to charm-quarks have produced only weak limits to date.
Again, one of the reasons for this poor performance is the \gls{sm} the rate for $h \to b\bar{b}$ is about 20 times larger than the rate for $h \to c\bar{c}$. 
The standard approach relies on tagging the flavour of the jets, which involves discriminating charm initiated jets from bottom jets, or vice versa.
The primary technique used currently in this case is Boosted Decision Trees, mainly structured as binary classification problem, where the community effort is devoted to the definition of ad-hoc flavour tagging through the use of the class imbalance techniques instead of general purpose ones~\cite{Aaboud_2018}.

It is natural to ask whether quantum computing algorithms could be used to support these complicated tasks. However, it is not evident where a quantum algorithm could provide a systematic advantage with respect to these classical approaches. Possible directions of research should answer the following questions: can we overcome the problem of lack of density or insufficiency of information for these problems? Can we better explore and analyse the feature space that describes those problems? Could \gls{qml} methods, which employ quantum models to encode input data into a high-dimensional Hilbert space and extract physical properties of interest from the quantum state, be an alternative approach to signal detection? A particularly intriguing direction for quantum approaches here could be the possibility of training directly on experimental data~\cite{Metodiev_2017} that can be directly analysed as quantum data.

Overall, in the absence of a clear hint of new physics in \gls{hep} experiments, a data-driven, model-agnostic search for rare signals has gained considerable interest. Anomaly detection, realized using unsupervised machine learning, is the most commonly used technique and will continuously becoming important in \gls{hep} analysis workflow. 
The feasibility of anomaly detection is investigated in~\cite{PhysRevD.105.095004} with \gls{vqa}-based \gls{qae}. With the benchmark process of $pp \to H \to t\bar{t}$ for signal, the \gls{qae} performance for anomaly detection has been compared with that from a classical autoencoder, showing a faster convergence in the quantum case. Recently, in~\cite{schuhmacher2023unravelling}
 the authors find that employing a \gls{qsvc} trained to identify the artificial anomalies, it is possible to identify realistic \gls{bsm} events with high accuracy. In parallel, they also explore the potential of quantum algorithms for improving the classification accuracy and provide plausible conditions for the best exploitation of this novel computational paradigm. Additionally, in~\cite{QADCERN}  the authors found evidence that quantum anomaly detection using a \gls{qsvm} could outperform the best classical counterpart.  In~\cite{bermot2023quantum} an \gls{aqgan} is introduced to identify anomalous events (\gls{bsm} particles). Interestingly, this model can achieve the same anomaly detection accuracy as its classical counterpart using ten times fewer training data points.

Overall, current quantum-classical hybrid \gls{qml} for rare signal extraction are  largely based on two algorithms: \gls{vqa}~\cite{VQA2021} and \gls{qsvm} with kernel method~\cite{havlivcek2019supervised,Schuld2019QML}.
The quantum kernel-based \gls{qsvm} has a potential for good trainability due to a convex cost-function landscape, and this property could be beneficial for the $100 \otimes 100$ challenge. It is however pointed out that the kernel function would exponentially concentrate to a fixed value with the number of qubits unless the $U({\boldsymbol x})$ is properly designed~\cite{thanasilp2022exponential}, analogously to the barren plateau in \gls{vqa}. 

\gls{vqa}-based \gls{qml} methods are generally known to be affected by the infamous barren plateau problem, where a non-convex landscape of cost function causes the gradients to vanish exponentially in the number of qubits,
as detailed in Sec.~\ref{subsec:qml_limitations}. 
With the $100 \otimes 100$ IBM challenge, overcoming barren plateaus may be critical for \gls{qml} applications to signal extraction. The approach based on so-called geometrical \gls{gqml}, that exploits prior knowledge to the problem, such as symmetry presented in the data at hand, will be promising for applications to \gls{hep} data analysis. However, experimental data are the result of a complex convoluted effect given by different layers of interaction, from parton shower to detector effects. This would eventually destroy any desirable symmetry of the data. Alternatively, quantum models in an overparameterized regime may have a desirable cost landscapes. This motivates exploring \gls{gqml} models and/or overparameterization in a realistic \gls{hep} data analysis flow. We should also pursue how efficiently a \gls{qml} model can generalize to unseen test data with fewer trainable parameters or less training data, and also consider the possibility of re-using well known techniques from classic machine learning, like ensemble, where, for instance, the effect of noise could be mediated by the structure of the algorithm~\cite{incudini2023resource}.

\subsubsection{Pattern Recognition Tasks: Reconstructing Particle Trajectories and Particle Jets}
Multiple steps in the experiments data processing chains can be collected into the general category of pattern recognition or the problem, given a certain number of measurements of an object (such as the raw energy measured by the sensors in a detector, or its spatial coordinates), of associating them to a specific instance: for example a particle trajectory, a particle type, the particle jet that originates from the hadronization of a specific parton (jet).   
In \gls{hep}, this problem has high dimensionality, since the detector sensors are arranged in highly granular structures, the objects represent physics properties and the object classes are typically exclusives (an energy deposition belongs to one and only one trajectory). 
Two examples are indeed represented by the reconstruction of charged particles and the reconstruction of jets, together with the identification of their properties.

The reconstruction of charged particle trajectories, \textit{tracking}, is an essential ingredient in event reconstruction for \gls{hep}. Particle track candidates are built from space points corresponding to energy deposits left by charged particles - or \textit{hits} - as they traverse the sensitive detector material. The track parameters (e.g. position and curvature) hereafter computed are used in subsequent processing steps throughout the reconstruction and analysis of data to compute physics observables.

In collider particle physics, a jet is a collection of stable particles collimated into a
roughly cone-shaped region. Jets arise from the fragmentation of quarks and gluons
produced in high-energy collisions. During the collision, the \gls{qcd} confinement the
quarks and gluons are subjected to is broken, yielding a spray of color-neutral
particles that can be experimentally measured in particle detectors. Jets have played
and are playing a fundamental role in collider physics. Events with three jets in $e^+ e^-$
collisions demonstrated the existence of the gluon. Nowadays jets produced by the
fragmentation of heavy quarks, namely $b$ and $c$ quarks are crucial for several studies
in particular to determine the Higgs boson couplings. In the latest years tools have
been developed to disentangle different kinds of jets.

\paragraph{Track Reconstruction}
Several current and future \gls{hep} experiments will explore high intensity scenarios going to extreme regimes with thousand of charged particles crossing a square centimeter of sensitive detector. Furthermore, depending on the process under study and the detector layout, each track can consist of a variable number of measurements. The multiplicity of possible track candidates from the input space-points scales quadratically or cubically with the number of hits. Therefore tight selections on the input space-points are required in order to narrow down the search space. Nevertheless, track reconstruction is one of the largest users of CPU time in \gls{hep} experiments, strongly motivating the R\&D of novel approaches.

Several approaches have been proposed to address the tracking problem and can be roughly divided into global and local approaches. Global tracking methods approach track reconstruction as a clustering problem, thus considering all the space-points at once, whereas local tracking methods generally consist of a series of steps executed sequentially. Several studies have been performed for both global~\cite{Funcke:2022dws,Crippa:2023ieq} and the local~\cite{Magano:2021jzd} methods, finding a potential reduction of computational complexity for the latter. 

First proposals to solve the particle track reconstruction problem on a quantum computer focused on converting the problem to a \gls{qubo} problem~\cite{bapst_pattern_2019, zlokapa2021}. This way, one can group two (doublets) or three (triplets) hits from consecutive detector layers and binary values represent if a given doublet or triplet corresponds to a particle track. 
There have been several proposals in the literature on how to determine the coefficients of the \gls{qubo} either based on geometry or impact on the overall energy of the \gls{qubo}~\cite{Funcke:2022dws, schwaegerl2023particle, Crippa:2023ieq}. In its most general form, one can write such a \gls{qubo} Hamiltonian as,
\[H=\sum_{i < j}^{N} J_{i,j}T_iT_j + \sum_{i=1}^{N}h_iT_i, \]
where in the case of triplets, $T_i$ represents $i^{th}$ triplet and $T$ can be mapped to the Pauli-Z operator on the $i^{th}$ qubit. $J_{ij}$ is the coupling coefficient between $i^{th}$ and $j^{th}$ triplets and $h_i$ determines the strength of the field on the $i^{th}$ triplet. Then, it is possible to use algorithms such as \gls{qaoa}, \gls{vqe} or \gls{hhl} to find the ground state of the Hamiltonian, which corresponds to the desired solution. Although such a \gls{qubo} Hamiltonian is sparse in general, it consists of at least $\mathcal{O}(10k)$ sites for a real world problem, or $\mathcal{O}(500)$ in more favourable scenarios with smaller occupancies such as the LHCb Vertex Locator. The limited number of qubits available currently restricts the Hamiltonian to $\mathcal{O}(1)$ sites and therefore strategies to partition the Hamiltonian to many smaller pieces are needed.

Recently, Ducket et al. proposed a method to solve the triplet classification using a \gls{qsvm} based approach~\cite{duckett_reconstructing_2022}. In this method, spatial coordinates of each hit from the triplet are encoded to quantum states that result in a 9-qubit circuit. Quantum kernel methods promise an advantage for datasets with many features, therefore a triplet based approach might not provide an advantage. However, this method may outperform a classical kernel method in cases where considering higher number of hits are useful.

Classical \gls{gnn} methods were shown to have linear scaling with respect to the number of input space-points, which makes them a strong candidate for future implementations of particle track reconstruction algorithms~\cite{ju_performance_2021}. Although there is no formal proof that this scaling is linear, the empirical evidence suggests so. It is likely that this improvement comes from the parallelisation capacity of \gls{gpu}s. This means that there is still a need for large \gls{gpu} clusters. A quantum advantage could be achieved if \gls{gnn}s with similar characteristics can be implemented on Quantum Computers. Recently, it was shown that a \gls{qgnn} approach is possible and it can perform similar to the classical equivalent for up-to 16 qubits~\cite{tuysuz21, QGNN-embedding}. However, understanding if this can be realized at large scale requires a larger number of qubits. 

The availability of a $100 \otimes 100$ device would enable the study of larger local Hamiltonians with 100 sites and give researchers a tool to investigate if \gls{qubo} based approaches are viable. Similarly, such a device would allow us to implement \gls{qgnn}s of sizes comparable to the classical state-of-the-art models. 

Regarding local methods, although a full analysis chain is presently unreachable due to hardware limitations, we can nevertheless consider a complexity analysis to illuminate the general evolution of the classical and quantum approaches to the problem. It is not clear, in particular, whether all steps in the track or, in general, object reconstruction may benefit from a quantum algorithmic approach.
This is the procedure originally followed, for instance, by Wei et al.~\cite{Wei2020}, which have estimated the classical and quantum computational scaling of a well-known (albeit unused) jet clustering algorithm. However, since this algorithm is not the current standard used at the LHC it is much more informative to estimate the complexity of a current choice, namely the \gls{ctf} algorithm~\cite{JINST}, which is the tracking algorithm used by the CMS collaboration~\footnote{New versions have been published~\cite{Bocci2020}, but the general analysis should hold also in this case.}. The underlying structure of the \gls{ctf}, the combinatorial Kalman filter~\cite{KalmanFilterFruhwirth}, is used by several current track reconstruction algorithms~\cite{Sguazzoni2016,ATLAS,Braun2018} and the analysis can easily generalized to most presently available algorithms. This program has been followed by in~\cite{Magano:2021jzd} using the algorithm as it is described in~\cite{JINST}. The conclusion is that it is possible to reconstruct the same tracks (up to bounded-error probability) with lower quantum complexity by an adequate use of quantum search routines.

A $100 \otimes 100$ machine may allow for some progress along the lines defined in~\cite{Magano:2021jzd}, although it is necessary to investigate the number of qubits which can effectively be used for the implementation of the program. Moreover, since~\cite{Magano:2021jzd} is applicable to a hybrid classical/quantum approach it is possible to implement the program according to the available resources. In any case, it is important to bear in mind that in the short-term track reconstruction algorithms will be quite limited by the input size, and investment in \gls{qaoa} or in jet clustering may be more rewarding.

\paragraph{Jet Reconstruction and Identification}
\label{subsubsection:Jet reconstruction and identification}

Jet clustering algorithms aim at estimating the kinematics of the particle that initiated the jet. Usually, these algorithms are based on clustering schemes, which combines the observed particles into a jet for further study.

Clustering algorithms have different properties and characteristics that can make them more appropriate for a particular task, such as the extraction of observables or as a tool to extract specific properties of the final state. An essential property of an optimal jet clustering algorithm is \gls{irc} safety. An observable is \gls{irc} if it remains unchanged in the limit of a collinear splitting or the emission of an infinitely soft (low momentum) particle.

Two main approaches have been pursued in clustering particles into a jet: cone and sequential recombination schemes. The first approach aims to find regions with a high-energy flow and thus define rigid conic boundaries. In sequential recombination algorithms, particles are clustered locally using a distance metric.

Jet clustering algorithms can be computationally expensive, as the execution time scales polynomially with the number of particles to cluster. Speedups can be achieved by considering the clustering problem from a geometrical point of view instead of combinatorially. In this way, sequential recombination algorithms can be executed in $\mathcal{O}(N^2$) or even $\mathcal{O}(N \ln N$) complexity rather than $\mathcal{O}(N^3$). Cone algorithms could be implemented exactly (and therefore made \gls{irc} safe) with $\mathcal{O}(N^{2}\ln N$) rather than the expected $\mathcal{O}(N2^{N}$) complexity. 

Quantum-assisted algorithms have been explored to reduce the computational overhead of these clustering routines. The first application of quantum-assisted algorithms to the task of clustering particles into a jet was introduced in~\cite{Wei2020}. Two clustering techniques were put in place for the particular case of electron-positron collisions and inspired by the calculation of \textit{thrust}~\cite{BRANDT196457,PhysRevLett.39.1587}, an event shape quantity that allows for the partition of event particles into two hemisphere jets. The first approach targeted the universal quantum computing setting based on Grover's algorithm. In addition, a \gls{qubo} formulation for thrust was developed, suitable for quantum annealing. Classically, the calculation of thrust can be costly, scaling as $\mathcal{O}(N^3)$~\cite{YAMAMOTO1983597} for an event with $N$ particles, or using the improved method introduced in~\cite{Salam_2007}, as $\mathcal{O}(N^2 \log N)$. The thrust-based \gls{qubo} formulation was benchmarked in~\cite{DelgadoThaler2022}, using the D-Wave Advantage 1.1 QPU, and compared to classical \gls{qubo}-solving techniques such as simulated annealing and annealing optimization subroutines like reverse annealing. Results from these studies revealed the limitations of current quantum annealing devices in terms of connectivity. \gls{qubo} formulations involving many spin variables and all-to-all connectivity, like the thrust problem, perform poorly on currently available quantum annealers. An extension to the \gls{qubo} formulation for thrust calculation was presented in~\cite{pires2020adiabatic}, based on the angular distance between two particles in a given event and penalizing the assignment of two particles located on the same hemisphere of the partition. Results from the hardware deployment of these studies were limited to a low number of annealing runs due to limited access to the QPU.

Algorithms based on digital quantum computing have also been proposed; however, the algorithms in~\cite{pires2021digital, Wei2020} are not suitable for implementation on noisy devices due to the need for a QRAM-like architecture to access particle information in parallel. Another promising study~\cite{deLejarza:2022bwc} deals with the quantum version of three clustering algorithms found in the classical literature: \textit{k-means}~\cite{MacQueen1967} (a quantum version of this algorithm is used in Ref.~\cite{pires2021digital}), \textit{affinity propagation}~\cite{Brendan2007}, and the \textit{$k_T$} jet clustering algorithm~\cite{PhysRevD.48.3160}. Two quantum subroutines are introduced: the first computes the Minkowski distance between particles, and the second tracks the maximum in a long-tailed distribution. For both these subroutines, the authors prove polynomial speedups as compared to well-known classical algorithms.  The quantum algorithms were applied to simulated data for a typical LHC collision setting and obtained efficiencies comparable to their classical counterparts. In particular, the quantum-\textit{$k_T$} version is a conceptually more straightforward algorithm with a similar execution time compared to subroutines in the FastJet library~\cite{Fastjet}. 

Jet tagging, the identification of the flavor of the quark that originated the jet, is another aspect of jets physics that experimental physicists are continuously improving. For example, in the determination of the Higgs boson couplings to $b$ and $c$ quarks, the jet tagging efficiency and purity determine the actual size of the dataset useful for the measurements and, therefore, their accuracy. Jet tagging is based on global jet characteristics and on each jet's particle properties. In principle, it therefore requires a large number of features which means a high dimension dataset. 
The study reported in~\cite{btag} limited the data representation to a few properties to cope with the low number of qubits available and short circuit depth. As already mentioned in the introduction of this section, this is the approach often used in experimental \gls{hep}, and therefore the performance of the QC algorithms is by definition limited and the comparison with classical methods is performed adopting the same data-set dimension. 
Two different feature encodings have been tested: the angle encoding is used when a two feature data-set is used while for the 16 features the amplitude encoding is exploited.
Even though the exercise is quite simple, it showed that in the training phase, the \gls{qml} method reaches optimal performance with a lower number of events with respect to the classical ones. The limited access to the hardware resources did not allow an extensive study of the noise impact which was evaluated only for the 2 qubits case.
This study could largely benefit from much more powerful hardware, in particular the $100 \otimes 100$ IBM hardware.
Instead of re-proposing the same exercise, it would be possible to design a new circuit where the entanglement entropy can play an important role. In fact, jet tagging features correlation is considered as classical correlation while in \gls{qml} these can be understood and included in the circuit optimization improving the classification performance. 
A further step forward could be to perform the jet classification study on data obtained in proton-proton collisions and in Monte Carlo simulated events. Collider data may exhibit quantum characteristics,   not visible in simulation. That could happen due to the limited knowledge of jets formation and evolution which is regulated by the non-perturbative \gls{qcd} and described only by models in the simulation. Such effects, if there, are currently  absorbed by the systematic errors in the jet reconstruction quantities.

\subsubsection{Interpretable Models and Inference}
 
 In this section we review the use of quantum models as inference tools to extract the characteristic properties of a dataset in \gls{hep}. We give two examples of such models: characterising the non-perturbative structure of hadrons through \gls{pdfs}, and estimating the Wilson coefficients of \gls{efts} and their correlations. We emphasise the potential of these tools to enable precision modelling of physical phenomena and provide a first step towards being able to bridge the fields of quantum computing and quantum information in extracting quantum descriptors of \gls{hep} processes from models learned from data. 

In high-energy physics, perturbation theory is used to calculate particle interactions at high energies~\cite{Gross:1973id}. These perturbative methods allow for the calculation of scattering amplitudes as a series expansion in powers of the coupling constant. However, as the energy of the interaction decreases, the coupling constant becomes large and the perturbative expansion breaks down~\cite{Gross:1973ju}. This results in a non-perturbative regime where the underlying physical processes are not easily calculable, and must instead be obtained through experimental measurements or numerical simulations~\cite{Shifman:1978bx}. Characteristic functions that capture the essential features of the underlying physical process can be employed to represent the relative probability of a particular physical process as a sum of simpler, more tractable functions. The choice of basis functions and coefficients is critical in constructing an accurate representation of the underlying physical process as they must be able to accurately characterize the process in both the perturbative and non-perturbative regimes, as well as any additional physical constraints that may be present. 

\gls{pdfs} are an example of non-perturbative effects that are necessarily characterised by such approximations from data~\cite{collins_2011}. \gls{pdfs} describe the probability distribution of the momentum fraction carried by the quarks and gluons inside a proton. The need for \gls{pdfs} arises from the fact that the proton is a composite particle made up of quarks and gluons that are constantly interacting, which makes it impossible to calculate the momentum distribution of these partons using perturbative methods alone. Nonetheless, there are known constraints on the form of \gls{pdfs} that can be derived from the fundamental principles of quantum chromodynamics (\gls{qcd}), and their predictions are highly constrained by experiment.

The accurate estimate of \gls{pdfs} is vital to all measurements in experimental collider physics, as they are used to predict the rates and distributions of processes~\cite{practical_collider_physics}. 
Uncertainties arise from the limited precision of how experimental data is used to constrain the \gls{pdfs}, as well as from theoretical uncertainties in how to extract perturbative estimates from the fitted \gls{pdfs}.
Quantum computing affords us new avenues to address both of these shortcomings by providing characteristic functions that may better represent the nature of the process they are used to represent.

A recent study investigating an approach based on the use of a \gls{pqc} was explored for estimating the functional form of \gls{pdfs}~\cite{Perez-Salinas:2020nem} from data. The \gls{pqc} approach aims to find an ansatz for representing the \gls{pdfs} as a \gls{pqc}, the parameters of which are estimated using a classical optimization algorithm to minimise the difference between the predicted and experimental data. This is a promising avenue for leveraging the expressive power of \gls{pqc}s to efficiently learn solutions to classically intractable problems. Preliminary results using the \gls{pqc} approach are encouraging, showing good agreement with existing \gls{pdf} fits obtained through classical optimization techniques. This represents an exciting first step towards using quantum algorithms for \gls{pdf} estimation and highlights the potential of quantum computing for solving problems in high-energy physics. However significant work is still needed to leverage the quantum nature of the problem. In this construction, the \gls{pdfs} being estimated are still classical approximations to an inherently quantum system and as such the possible advantages of such a methodology are purely computational. It is foreseeable that quantum functions that characterise classically intractable processes such as these are possible in a way that, although simplified with respect to a fully numerical model (e.g. from lattice \gls{qcd}), would give a notable improvement over classical models when compared with data.

In contrast to experimental measurements in which the exact prediction of a given standard model process is computed, \gls{efts} provide a framework for modelling complex physical processes in terms of a hierarchy of simplified interactions, characterised by a set of Wilson coefficients~\cite{Lang:2021hnd}.
These coefficients represent the coupling strengths of various operators that encode the effects of high-energy physics and can be determined through a process of matching with experimentally measured observables.
While the precise values of the Wilson coefficients cannot be computed exactly, they can be approximated through a process of functional approximation, in which an ansatz is made for the form of the \gls{eft} such that the coefficients can be estimated from experimental data.
This \gls{eft} approach is similar in nature to the fitting of \gls{pdfs}, as both involve characterising complex physical phenomena in terms of a simplified set of paremetrised functions. 
In a recent study~\cite{Criado:2022aoo}, researchers proposed a new method for estimating Wilson coefficients using a quantum computer. The method involves using a quantum computer to encode the \gls{eft} predictions and experimental data into a \gls{qubo}~\cite{QUBO} optimization problem, with coefficients of the cost function determined by a Hamiltonian representation of a set of given coefficients. The \gls{qubo} problem is then solved on a quantum annealer to obtain the best-fit values of the Wilson coefficients. A primary goal of this method is that the optimization of the problem on a quantum computer can provide a more efficient and accurate way of estimating the Wilson coefficients than is possible with classical methods. 
The Hamiltonian is constructed using the parametrisation of an effective field theory approach, which allows for a systematic expansion in powers of the inverse of the mass scale of the new physics being probed.
The Standard Model Effective Field Theory (SMEFT) framework~\cite{SMEFT} used in this example contains a large number of parameters, making it challenging to extract information about the underlying physics. 
In this case the predictions for a reduced set of parameters are computed classically and only the relationship between the coefficients and the measurements modelled in the hamiltonian, not the dynamics of the \gls{eft} operators themselves.
Through a careful understanding of the Hamiltonian representation of this process and its solution using \gls{qml}, it might be possible to reduce the number of parameters needed to describe the system by identifying a smaller set of effective parameters that capture the essential physics and to identify correlations between different observables and effective couplings.

Whilst much of the development of quantum algorithms and in particular \gls{qml} is focused on identifying and solving computational bottlenecks present in traditional methodologies, the goals identified in these selected benchmark applications are those that leverage properties of \gls{qml} models to better interpret data from experimental high-energy physics in new ways. 
Several studies have begun towards this goal, however these initial steps provide only hints to which a complete understanding of how \gls{qml} can be used to leverage a quantum interpretation of the information contained in data from experimental particle physics.
 
 \subsubsection{Generative Models for Simulation}

 Other natural applications of generative modelling, are detector simulation and event generation.
Monte Carlo simulation of collider detector events is one of the most computing expensive tasks within the experiments data processing chain. Recent estimates suggest more than 50\% of the LHC  computing grid (WLCG) is spent on simulation tasks directly or on the simulated data reconstruction, i.e. the extraction of high-level features from simulated data~\cite{HSF}. The next generation detectors for the  High Luminosity LHC and future colliders, with their larger sizes, higher sensor granularity and increased complexity, will be even more demanding in terms of computing resources for data simulation and reconstruction~\cite{schmidt2016high}. 
This fact has sparked, over the years, intense research on alternative approaches, generally designated as fast simulation strategies in contrast to highly accurate Monte Carlo-based simulation. 

Fast simulation, typically trading some level of accuracy for speed, relies on parametric modelling of detectors response~\cite{ovyn2009delphes} or, more recently, on deep generative models~\cite{khattak2022fast,krause2021caloflow,buhmann2021getting} that learn multi-dimensional, conditional probability distributions.
In most cases, the focus is on the detector response itself: the deep generative models are trained to reproduce the detector output which is then processed in the same fashion as Monte Carlo simulated data. This approach can produce very realistic output, both in terms of quality of the individual events and in terms of sample diversity. In other cases direct generation of high level features, typically used at the analysis level, is preferred, thus skipping the entire reconstruction process~\cite{hashemi2019lhc}. This approach has the advantage of being computationally lightweight and flexible since the deep learning models learn directly the particles features and correlations in the final state of interest, taking into account all experimental effects. Its main limitation is the fact the output is inherently analysis specific, and cannot be used outside the scope it was initially designed for.

At the same time, several studies have started investigating quantum (or hybrid quantum-classical) implementations of generative models. A few examples are described in~\cite{arxiv.2203.08805} and references therein. In most cases quantum architectures inspired by classical models have been studied: for example, implementations of \gls{qgan} or \gls{qae}.  Particularly interesting is the case of \gls{qcbm}~\cite{qcbm} which instead are quantum generative models that do not have a classical counterpart and leverage the Born measurement rule during the sampling process. 
As in the classical domain, quantum architectures have been used to address two main types of applications: detector output simulation and final state generation. In both cases, but in particular for detector output simulation, the main limitation of the current models lies in the dimensionality of the simulated output. Realistic applications to particle physics detectors require generative models to learn distributions whose size scales with the number of detector sensors. 

 Current models can generate accurate simulations for very small (10-sized) setups, using one qubit to represent a detector sensor. Typically, reversible data compression techniques, such as auto-encoders,  can be used to bring down the original simulation to a size that is manageable by the quantum system: the classical encoder network produced a reduced latent representation which is then learned by the quantum generative model. A classical decoder network is then used to transform the synthetic output from the latent dimension to the original one. The expected advantage of using a quantum algorithm in this task would come from a more accurate and generalisable learning of the latent representation. It is clear, however that a most interesting development in this direction would require reducing the weight of the classical data dimensionality reduction step with respect to the quantum algorithm. In this context, a $100 \otimes 100$ machine would enable the simulation of far more realistic use cases.
 
Aside from the sheer detector size, the need for discretization also affects how realistic quantum generative models-based simulations can be. In most cases, our detectors produce continuous features, while qubits  naturally map to discreet quantities so the size of the qubit system can have an impact on the detector simulation resolution. 
These same problems can affect the direct generation of high-level features, albeit at a different scale: in this case quadri-momenta (and possibly angular correlations) of particles produced in scattering or decay processes are in the range of a few tens, instead of a few thousands, making the problem much more manageable on near term quantum systems. In this case, extreme care should be put into correctly describing cross-correlations among particles, thus good connectivity and the possibility to reproduce complex entanglement patterns over multiple qubits become essential.

\section{Algorithms, Methods and Limitations\label{sec:algs}}

\subsection{Quantum Algorithms for Quantum Dynamics}
\label{subsubsec_algo_dyn}

The development of quantum algorithms for the simulation of quantum dynamics is a very active field of research, with potential applications covering a broad spectrum across the physical sciences~\cite{Tacchino2020,Miessen2023}.
A plethora of powerful methods has been developed over the past years, which can generally be classified as either decomposition-based or variational in nature~\cite{Miessen2023}. Techniques belonging to the former category aim at realising a target unitary evolution $U(t)=e^{-iHt}$ through a decomposition into elementary quantum logic operations. This approach typically yields rigorous scaling laws, a priori error bounds and, most importantly, provides a systematic way of exchanging resources (i.e., circuit depth, gate counts, ancilla qubits) for accuracy. Examples of decomposition methods include product formulas (see below), linear combinations of unitaries~\cite{berry2015simulating}, quantum signal processing~\cite{low2017} and qubitization~\cite{Low2019hamiltonian}. On the other hand, variational strategies address the task of approximating $U(t)$ by resorting to parametrized quantum circuits, for example implementing time-dependent ansatzes or learning effective partial representations of the dynamics. This often reduces the circuit complexity compared to decomposition methods, thus lowering the experimental requirements for implementations on current noisy quantum processors. However, such an advantage comes at the cost of some classical overhead (e.g., optimization, additional measurements) and within a more heuristic framework where accuracy guarantees are harder to obtain. Both decomposition and variational algorithms have been applied for specific dynamical studies in \gls{lgt} on quantum computers~\cite{Mathis2020,nagano2023}.

\subsubsection{Product Formulas} 
Among decomposition methods, product formulas represent the simplest and most widely adopted paradigm~\cite{Lloyd96,chiesa_quantum_2019,Kim2022Scalable}. In their basic implementation, these rely on the general Trotter approximation rule~\cite{trotter1959product}
\begin{equation} \label{eq:trotter}
    e^{-i\mathcal{H}t} = \lim_{n\to\infty} \bigg(\prod_i^M e^{-i\mathcal{H}_it/n} \bigg)^n 
\end{equation}
where $\mathcal{H}=\sum_i^M \mathcal{H}_i$. At first order and for every finite choice of $n$, one has 
\begin{equation}
    e^{-i\mathcal{H}t}\simeq \biggl(\prod_i^M e^{-i\mathcal{H}_it/n} \biggr)^n + \mathcal{O}\bigl(\sum^M_{i>j} \lVert [\mathcal{H}_i, \mathcal{H}_j)]\rVert t^2 / n \bigr),
\end{equation} 
i.e., the decomposition error amounts to $\mathcal{O} ( M^2 t^2/n )$. This may be systematically reduced either by choosing a larger $n$ or by employing higher-order \gls{pf}, for which the error becomes $\mathcal{O} ((M \tau)^{2k+1} / n^{2k})$ at order $2k$ ($k \geq 1)$ (see also Sec.~\ref{subsubsec:limitsQD}). In both cases, better theoretical accuracy is obtained in return for an increased gate count. Further improvements are possible, based, e.g., on randomization and adaptive techniques~\cite{zhang2012randomized,Childs2019fasterquantum,campbell2019,Zhang2023Low} or on the use of linear combinations of \gls{pf} (multi-product formulas)~\cite{chin2010multi,Vazquez2022}, which can reduce Trotter errors. 
Importantly, \glspl{pf} can also be employed for simulating time-dependent Hamiltonians~\cite{wiebe2010higher,poulin2011quantum,watkins2022timedependent}.

\subsubsection{Variational Approaches} 
Parametrized quantum circuits can be used to tackle quantum dynamical problems by either resorting to well-established variational principles or by recasting them as optimization tasks. In the first case~\cite{Yuan2019}, one builds a time-dependent wavefunction ansatz spanning a suitable manifold in the Hilbert space of the target system and propagates the parameters by solving a classical \gls{eom}. For sufficiently well-behaved dynamics, the trajectory of a specific quantum state in time can be approximated with a number of parameters that is significantly smaller than the dimension of the full space. This in principle results in a simulation whose cost in terms of quantum resources -- and specifically circuit depth -- is constant, or only increases moderately, with time (in contrast to, e.g., \glspl{pf}). As an example, for an ansatz $\ket{\Phi(\theta(t))} \equiv \ket{\Phi}$ evolving under the action of a Hamiltonian $H$, the application of McLachlan’s variational principle leads to a set of differential equations for the parameters of the form~\cite{Yuan2019}
\begin{equation}
    \mathcal{M} \dot{\bm{\theta}} = \bm{\mathcal{V}} 
\label{eq:McLachlanEOM}
\end{equation}
where
\begin{equation}
    \mathcal{M} = \Re \biggl( \frac{\partial \bra{\Phi}}{\partial \theta_i} \frac{\partial \ket{\Phi}}{\partial \theta_j} + \frac{\partial \bra{\Phi}}{\partial \theta_i} \ket{\Phi} \frac{\partial \bra{\Phi}}{\partial \theta_j} \ket{\Phi} \biggr)
\end{equation}
and
\begin{equation}
    \bm{\mathcal{V}} = \Im \biggl( \frac{\partial \bra{\Phi}}{\partial \theta_i} H \ket{\Phi} - \frac{\partial \bra{\Phi}}{\partial \theta_i} \ket{\Phi} \braket{\Phi | H | \Phi} \biggr) \ .
\end{equation}
The matrix elements of $\mathcal{M}$ and $\bm{\mathcal{V}}$ have to be evaluated through measurements on the quantum processor where the ansatz is prepared, while Eq.~\eqref{eq:McLachlanEOM} is integrated classically. 
Two main versions of the \gls{vte} algorithms have been devised and applied in quantum simulations: the \gls{varQTE}  algorithm for real-time propagation, and the \gls{varQITE} algorithms for `dynamical' ground state preparation (for a review see~\cite{Miessen2023}). 

This \gls{vte} algorithm is particularly attractive in those cases where a direct decomposition of unitary Hamiltonian evolution becomes quickly demanding with growing system size, such as, e.g., in first quantization~\cite{ollitrault2022quantum} or when dealing with fermionic/bosonic degrees of freedom~\cite{macridin2018,endo_calculation_2020,miessen2021quantum,LibbiGreen2022} and gauge fields~\cite{Mathis2020, Mazzola2021}. 
In practice, one crucial ingredient is the choice of the ansatz, which ideally should incorporate physical intuition (e.g., respecting symmetries and/or conservation laws) and good mathematical properties (e.g., concerning the form of the tangent space associated with the parametrized manifold along time-evolution paths). While several promising strategies for ansatz construction have been proposed, including adaptive 
ones~\cite{gomes2023computing}, it remains challenging, in general, to correlate in a systematic way ansatz expressivity with simulation accuracy and performances, if not with a posteriori error bounds~\cite{zoufal2021error}. The application of \gls{vte} is also limited by the high numerical sensitivity associated with the solution of Eq.~\eqref{eq:McLachlanEOM} via matrix inversion and by the large number of measurements required to construct $\mathcal{M}$ and $\bm{\mathcal{V}}$~\cite{miessen2021quantum, zoufal2021error}. 

In parallel to the standard \gls{vte} algorithm, numerous other approaches are being explored. For instance, variational quantum methods have been employed to learn a (partial) diagonalization of the short-time evolution of a system~\cite{heya2019subspace, cirstoiu2020variational, gibbs2021long, gibbs2022dynamical} and to compress the circuits required to implement a short time-step on a given state~\cite{otten2019noise, Lin2021Real, barison2021efficient, benedetti2020hardware}. Additional proposals aimed at implementing near-term quantum simulations include: quantum-assisted methods which perform all necessary quantum measurements at the start of the algorithm instead of employing a classical-quantum feedback loop~\cite{bharti2020quantum,lau2021quantum,haug2020generalized}, methods based around Cartan decompositions~\cite{Kokcu2022Fixed, steckmann2021simulating} and approaches using Krylov theory~\cite{jamet2021krylov}.  

\subsubsection{Algorithmic Limitations\label{subsubsec:limitsQD}}  
Of the two approaches for performing time-evolution dynamics, it is considerably more straightforward to characterise the (near-term) simulation errors associated with Trotterisation-based methods.
For a fixed total time $T$, the discretization in $n$ time-intervals ($dt=T/n$) of the time-evolution operator according to Eq.~\eqref{eq:trotter} (i.e., using first-order Trotter expansion) will lead to a residual error $\epsilon$ of $\mathcal{O}(\alpha_{\text{c}} (T/n)^{p+1})$ with $p=1$ where $\alpha_{\text{c}}=\sum_{i,j} || \mathcal{H}_j,[\mathcal{H}_j,\mathcal{H}_i]||$ (see~\cite{PhysRevX.11.011020}). 
This implies that one would require $n \times \mathcal{O}(M \alpha_{\text{comm}} (T/n)^{2})$ gates to achieve the desired accuracy, where $M$ is the number of Pauli strings building up the system Hamiltonian.
On the other hand, a practical implementation of the Trotter expansion in near-term, noisy, quantum computers will need to face the additional errors arising from the gate infidelities. Assuming only errors induced by the 2-qubit \gls{cnot} operations, $\epsilon_{\text{2g}}$, the overall Trotter error will scale as
\begin{equation}
    \epsilon_{\text{Trotter}} \sim \mathcal{O} (\alpha_{\text{c}} (T/n)^{2} + (n_0 \epsilon_{\text{2g}})^n )
\end{equation}
where $n_0$ is the number of 2-qubit operations required for the circuit implementation of the operator $e^{i \mathcal{H} t}$, for fixed $t$.
We therefore conclude that for the Trotter formula there must be an optimal value of the the discretization variable $n$, that we name $n^*$, which minimizes the overall error.

In the case of the \gls{vte} algorithm, the quantum circuit is of a constant depth, while the number of gates required for its implementation depends on the number of degrees of freedom necessary to produce a suitable representation of the subspace that spans the dynamics of the system. Assuming the knowledge of a variational form that can be systematically improved by adding circuit layers, $L$, one can - in principle - achieve a desired accuracy as a function of the circuit depth. 
In the case of \gls{lgt}, one could for instance employ a recently proposed Hamiltonian inspired variational Ans\"atze~\cite{Mazzola2021}, which has the advantage of combining a physical motivated variational circuit with the possibility of naturally enforce dynamical constrained, such has Gauss law. On the formal side, Ref.~\cite{zoufal2021error} has investigated error bounds associated with \gls{vte}. However, unfortunately, and in contrast to the Trotter expansion and alike, there appears to be no systematic ways to assess a priori the scaling of the variational error in \gls{vte} algorithm as a function of the total simulation time, $T$, or circuit layers, $L$. 
Preliminary studies~\cite{thesisNicola} showed that in the case of \gls{qed} calculations in \gls{1p1D} dimensions the number of Hamiltonian-inspired layers to reach a desired accuracy increases rapidly with the dimensionality of the problem, approaching the number of gates required to implement the Trotter formula already with 10 to 15 sites. 
Finally, it is also important to mention that the quality of the \gls{vte} approach depends on the accuracy in the solution of the system of linear differential equations in Eq.~\eqref{eq:McLachlanEOM}. 

\subsubsection{Near-term Applications} 
We would like to conclude this section with a rough estimation of the resources needed for the quantum simulations. For the \gls{qed} static study we take ground state properties as our main target (e.g., the important phase structure for which high precision is not necessarily required). Also for \gls{qed} dynamics we are interested in the qualitative behaviour of scattering particles.    
Therefore the entries presented below concern only the number of qubits and layers required (within the $100 \otimes 100$ challenge constraints) and do not refer to the precision of the solutions or the algorithms run times.

\begin{table*}
\caption{ 
For \gls{2p1D}~QED, we consider a minimum linear lattice size, $L$, of 4 and a maximum of 8; this leads to 8 to 16 qubits to describe the fermionic d.o.f. and 10/15 gauge links, which leads - with truncation of the gauge fields $l=2,3$ - to 20/100 and 30/150 qubits, respectively. The final number of resources is reported in the table. For \gls{1p1D} \gls{qed} dynamics we give a suitable number of lattice points which allows us to study the time evolution of scattering particles;
For two flavor neutrinos we consider a direct mapping to qubits, the cost is based on a first order \gls{pf}; for the largest system with 40 neutrinos the \gls{cnot} count would still be $2340$ with depth $120$.
}
\label{table:resources}
\begin{tabular}{| c | c| c | c | c |} 
  \hline
  \textbf{Systems}                 & \textbf{Phys.~size (min/max)} & \textbf{No.~qubits  (min/max)} & \textbf{Alg.}           & \textbf{No. CNOT layers}\\ 
  \hline 
  \hline
  \gls{2p1D}~QED static            & 4x4/8x8 sites                 & 30/160                         & \gls{vqe}/\gls{varQITE} & $\sim 10/100$   \\
  \hline
  \gls{1p1D} QED dynamics          & 12/20 sites                   & 30/100                         & \gls{varQTE}/Trotter    & $20/100$  \\
  \hline
  Collective Neutrino Oscillations & 10/40 neutrinos               & 10/40                          & \gls{vte}/\gls{pf}      & $30/120$  \\
  \hline
\end{tabular}
\end{table*}

\subsection{Quantum Machine Learning}

\subsubsection{Opportunities for Quantum Advantage}
\gls{qml} is an area of particular interest in experimental particle physics encompassing many of the algorithms described in this section. In general terms, two main approaches have emerged in the development of quantum-enhanced machine learning: the role of the quantum computer as an accelerator of otherwise established classical learning methods and the design of genuinely quantum methods, which do not mimic classical algorithms. This first method includes the relatively straightforward application of \gls{qc} methods to speed up an otherwise computationally costly training method~\cite{neven2008training}. Such approaches include classes of methods, dubbed \textit{quantum linear-algebra based methods}, in which the principle goal is to represent high-dimensional data in states of just logarithmically many qubits. This approach may allow even exponential speed-ups but comes with numerous caveats~\cite{Aaronson:2015scy}, most notably requiring some means of generating the required data-bearing quantum states, which, if done naively, already nullifies any possible advantage. Solutions to this may exist e.g. by the use QRAM~\cite{PhysRevLett.100.160501}, but in any case  these methods are mostly considered only in the context of large-scale fault-tolerant quantum computers.

In contrast, the design of genuinely quantum methods, may yet offer advantages based around the idea of parameterised quantum circuits (\gls{pqc}-based methods) as the key building block of the model. The basic examples here include the quantum support vector machine~\cite{havlivcek2019supervised} and the closely related quantum kernel methods~\cite{Schuld2019QML},  and, more generally, so called quantum neural network models. It is important to note that learning separations (so, provable exponential advantages) for learning using quantum models have already been proven in most learning settings ~\cite{liu2021rigorous}, subject to standard assumptions in complexity theory, and it can be shown~\cite{Gyurik:2022iqm} that these separations may be much more common when data is generated by a quantum process (under slightly stronger computational assumptions).

In general, the quantum model attains the form  $f_{\theta}(x)=Tr[\rho(x,\theta) O(x,\theta)]$, where the observable $O$ is most often fixed and independent from data ($x$) or trainable parameters ($\theta$), and $\rho(x,\theta)$ is prepared by applying a parametrized circuit on some fiducial state, e.g. $\rho(x,\theta) = U(x,\theta) |0\rangle \langle 0| U(x,\theta)$. In so-called linear models such as kernels and QSVMs, in contrast to data reuploading models~\cite{P_rez_Salinas_2020}, the state depends only on $x$, and this constitutes the loading of the data. Note, the targeted advantage in these settings is not in the dimensionality or number of data points, but rather in the \emph{quality} of learning that can be achieved. 

The mapping $x \mapsto \rho(x, \theta)$, a process which is typically independent of the setting of the $\theta$ parameters, constitutes the data loading, in which the key questions here are in finding a suitable mapping which will allow for a favourable data processing. Unlike in the case of quantum linear algebra approaches, the dimensionality of the state $\rho(x)$ is typically independent from the data dimensionality. In particular, as was proven in~\cite{P_rez_Salinas_2020}, already a single qubit line can express arbitrary multi-dimensional functions, given sufficient depth and data-re uploading. This is analogous how 1 hidden layer neural nets allow for functional universality~\cite{Cybenko1989ApproximationBS}, but nonetheless using multiple layers allows the more efficient access to useful function families. 
Using more qubits allows for more expressive function families at shallower circuit depths, and indeed qubit number scaling as a function of data dimension is necessary for any potential of a quantum advantage (as constant-sized quantum circuits are simulatable in polynomial time in the depth).

Indeed, the minimum is superlogarithmic scaling of the qubit numbers in the dimension of $x$, and linear scalings already can ensure the exponential cost of the classical simulation of the quantum model using best known classical algorithms.
This freedom also stymies any good approximations of how many qubits would be necessary to achieve good performance of quantum learning algorithms of this type; it is easy to construct models that are not classically simulatable, but at present it is not known how the increase of qubit number influences the quality of outcomes, and thus eventually outperform classical models.
In practice, it has been suggested that a linear scaling between qubits an input dimension may be a good starting point~\cite{Schuld2019QML}, however this necessitates the use of either classical dimensionality reduction techniques, or circuit cutting techniques~\cite{Marshall:2022jld} for any real-world applications in this field.

\subsubsection{Algorithmic Limitations\label{subsec:qml_limitations}}

A fundamental limitation to the scaling up most \gls{pqc}-based machine learning methods is the so-called barren plateau phenomenon, where the gradients~\cite{2018NatCo...9.4812M} of the cost function vanish exponentially with $n$. 
On such barren plateau landscapes, the cost function exponentially concentrates about its mean, leading to an exponentially narrow minima (a narrow gorge)~\cite{arrasmith2021equivalence}.  Hence, on a barren plateau, exponential precision is required to detect a cost minimizing direction and
therefore to navigate through the landscape.
Thus minimizing the cost typically requires an exponential number of shots, even if we use gradient-free~\cite{2020arXiv201112245A} or higher-derivative~\cite{cerezo2021higher} optimizers. While this phenomenon was originally identified in the context of variational quantum algorithms and quantum neural networks, it has recently been shown that  exponential concentration is also a barrier to the scalability of quantum generative modeling~\cite{rudolph2023trainability} and quantum kernel methods~\cite{thanasilp2022exponential}. 

A number of causes of barren plateaus have by now been identified, including using variational ansatze that are too expressive~\cite{2018NatCo...9.4812M,2022PRXQ....3a0313H, larocca2021diagnosing, tangpanitanon2020expressibility} or too entangling~\cite{PRXQuantum.2.040316, sharma2020trainability, patti2020entanglement}. However, even inexpressive and low-entangling \gls{qnn}s may exhibit barren plateaus if the cost function is `global'~\cite{2021NatCo..12.1791C}, i.e. relies on measuring global properties of the system, or if the training dataset is too random or entangled~\cite{PhysRevLett.126.190501, 2021arXiv211014753T, li2022concentration, 2022arXiv221101477L}. Finally, barren plateaus can be caused by quantum error processes washing out all landscape features, leading to noise-induced barren plateaus~\cite{2021NatCo..12.6961W,franca2020limitations}. 

Several methods to mitigate or avoid barren plateaus have been proposed. The simplest is perhaps to use a shallow ansatz along with a local cost function~\cite{2021NatCo..12.1791C, 2018NatCo...9.4812M}; however, it is questionable whether physically interesting and classically intractable problems can be solved within this regime. More promising is the ongoing search for problem-inspired ansatze~\cite{PhysRevX.11.041011,PhysRevLett.129.270501,2022arXiv220900292C,zhang2020toward,2020arXiv200802941W}, problem-inspired initialization strategies~\cite{2019arXiv190305076G}, pre-training strategies~\cite{Huggins_2019, dborin2022pretraining, rudolph2022synergy, rudolph2022decomposition, cheng2022clifford, niu2023warm} or layerwise learning~\cite{2020arXiv200614904S}. Of particular interest currently is the field of geometric quantum machine learning, which provides a group-theoretic strategy for building symmetries into \gls{qnn}s~\cite{larocca2022group, meyer2022exploiting, 2022arXiv220714413S, 2022arXiv221008566N}. 
In the context of \gls{lgt} simulations, this approach could be suitable since we can utilize the (local and global) gauge symmetry (see~\cite{Mazzola2021} for a gauge invariant construction of the ansatz, though it is not clear if this ansatz can mitigate the barren plateau problem).

Beyond barren plateaus there is a growing awareness of the problems induced by local minima~\cite{Bittel2021Training, you2021exponentially, rivera2021avoiding, anschuetz2022beyond}. Namely, it has been shown that quantum cost landscapes for a large class of problems can exhibit highly complex and non-convex landscapes that are resource intensive to optimize~\cite{Bittel2021Training, you2021exponentially, anschuetz2022beyond}. Thus constructing strategies to mitigate and avoid local minima~\cite{ rivera2021avoiding} is another important research direction to ensure the successful scaling up of hybrid variational quantum algorithms.

\subsubsection{Near-term Applications}
Given access to a noiseless 100-qubit system, assuming the capacity to train the model, it is in principle possible to tackle very high dimensional systems, and also learning problems where the underlying physics generating the data is very sophisticated. In essence, any system where a 100-qubit quantum simulation would be able to capture relevant physics, could in principle be captured by a 100-qubit learning model. In practice, as mentioned, this will only be possible if the \gls{pqc} architecture is carefully tailored to the learning task to enable the trainability of the system.

As shown in section \ref{subsect_Experiments} architectures inspired by the properties of classical deep neural networks have been successfully trained in the quantum context: quantum hierarchical classifiers, such as TTN and MERA, for example,  have been successfully trained to reproduce two-dimensional images representing the output of a \gls{hep} detector~\cite{rehm2023full, borras2023impact,chang2021dual}, quantum convolutions achieved optimal results for image analysis and image generation while mitigating the problem of barren plateaus~\cite{chang2022quantum}. A $100 \otimes 100$ machine could be used to understand to which extent a QCNN could reproduce the hierarchical learning of classical CNN, before incurring into limitations mentioned in the next subsection.

While a graph based interpretation of \gls{hep} data had been tested for a relatively small setup in the field of particle trajectory reconstruction, interesting quantum graph implementations have been proposed~\cite{mernyei2022equivariant} and could be tested in conjunction to point-cloud interpretation of \gls{hep} data for applications ranging from tracking to jet reconstruction and jet tagging to event generation of matrix element calculations. 
Quantum equivariant neural networks are also under study. Examples implementing spatial symmetries (rotations or reflections) have shown great potential on image related tasks and are being studied on \gls{hep} dataset as well. The case of physics symmetries, equally, if not more interesting, is also very promising, although for certain applications in classical data processing, a major challenge is represented by the difference existing between he original symmetries underlying the quantum process and the remnants accessible through measurements and observables.
An appropriate choice of loss functions and learning process will determine the task \gls{pqc}s can be trained for.

As explained throughout this paper, generative models are among the most powerful and versatile architectures that could be studied on a $100 \otimes 100$ machine: in particular, it should be possible to move from hybrid to fully quantum version of the more complex topologies such as \gls{qgan} or \gls{qae}. Designing a mechanism for efficiently reproducing attention on quantum states, could pave the way for the implementation of transformers, which are among the most powerful architectures existing today in the classical domain.
Similar considerations can be made in the choice of feature maps and kernels for kernel-based methods such as quantum support vector machines, which together with variational algorithms are used classification, clustering or anomaly detections problems, in the frameworks of both supervised or unsupervised learning.

\section{Conclusions and Outlook\label{sec:conclusion_outlook}}

In this paper we have described applications from experimental and theoretical high-energy physics where quantum computing has the potential to show a better performance than their classical counterparts. The selected applications were chosen also with respect to IBM's 100$\otimes$100 challenge and, where possible, a resource estimate was made. We note that the given applications are by no means complete and should serve as examples which are of very high interest for the high-energy physics community. We emphasize that this work should serve as an initial step by the present authors for exploring the potential of quantum computing for high-energy physics and we expect that the community of high-energy physicists working on this will substantially grow in the future. 

Concerning the quantum algorithms proposed for the applications outlined in the theory section (\ref{subsect_Theory}), we have identified quantum dynamics as one of the main targets because of its relevance in the field of \gls{hep}, e.g., in scattering phenomena, string breaking, quenching or dynamical properties of phase transitions. 
In fact, the exponentially growing costs of the corresponding classical approaches combined with  
the availability of well-tuned quantum algorithms make quantum computing a very promising tool for tackling problems in quantum dynamics. 
As already outlined in the theory section in table~\ref{table:resources}, such quantum dynamics applications are indeed compatible 
with the $100 \otimes 100$ challenge. 
Besides the dynamical aspects of theoretical \gls{hep} models applications we have also described static situations where quantum computing could lead to a better performance. These include abelian and non-abelian lattice gauge theories supplied with topological terms or non-zero fermion density or investigations of neutrino oscillations. While for these cases quantum computing has clearly an advantage over classical Markov Chain Monte Carlo methods it remains to be seen, whether it will have advantages over tensor network approaches, e.g. when taking the continuum limit or close to a phase transition. 

In this paper, we have identified and proposed concrete examples of low (1+1)D and (2+1)D theoretical models of \gls{hep} (and in particular lattice gauge theory) which are particularly hard classically due to the level of the entanglement produced, but still preserve a great physical relevance as prototypes for understanding fundamental dynamic but also static aspects of the laws of Nature.  
In the path towards large-scale simulations, we propose the development of hybrid quantum-classical algorithms, which can optimally leverage the advantages offered by the two complementary computational paradigms; for instance, the combination of \gls{tn} with quantum circuit representation of the system wave function can offer a unique opportunity for enabling the simulations of strongly entangled systems for longer time scales or close to phase transitions.

We consider the here proposed models as an intermediate step towards eventually reaching (3+1)D theories as actually needed  for studying the standard model of \gls{hep}. Besides the fact that the here considered lower dimensional models are of a high interest by themselves, we are convinced that investigating them with quantum computing can significantly help to develop algorithms and methods for studying their (3+1)D counterparts. And, it is a really fascinating outlook to explore phases of \gls{qcd} where no one has looked before such as the very eaerly universe or when the strenghts of a topological term becomes large. In addition, it would allow to study scattering phenomena in a fully non-perturbative fashion opening completely new insight to the physics of article collisons and shed light on the transition of confined phase of \gls{qcd} to the quark gluon plasma. 

A wide variety of \gls{qc} applications are anticipated in quantum simulations and \gls{hep} experimental workflow, as described in the earlier sections. Quantum simulations of simplified \gls{lgt}s in the Standard Model, such as \gls{2p1D} \gls{qed} or \gls{2p1D} SU(2) theory, are potential, well-motivated applications for near-term quantum computers. For the experimental side, \gls{qml} is a major technique to exploit quantum computing in the applications such as signal processing and detector reconstruction. 
However, as mentioned in Section~\ref{subsect_Experiments}, when considering \gls{qml} for processing classical experimental data, the data encoding into a quantum circuit is a big challenge, in particular for future colliders where an enormous amount of data will be produced. Moreover, the data encoding is also known to be one of the critical processes that cause barren plateau. 

This motivates us to explore the possibility of utilizing {\it quantum data} in the future as a promising route to directly exploit quantum properties encoded in quantum simulation and \gls{hep} experimental data.
From a theoretical perspective, understanding the power of quantum data for learning quantum states has received a lot of attention. There has been a sequence of works in tomography, wherein a learner is given copies to an unknown $n$-qubit quantum state $\rho$ and needs to learn $\rho$ well-enough (up to $\varepsilon$-trace distance); here the sample complexity was pinned down~\cite{haah2017sample,o2016efficient} to $\Theta(2^{2n}/\varepsilon^2)$. 
However, the exponential nature of learning an unknown quantum state is undesirable; there have been works that have looked at restricted classes of states and shown that they are learnable using polynomially many copies of the states, such as stabilizer states~\cite{montanaro2017learning}, Gibbs states of local Hamiltonians~\cite{anshu2021sample}, matrix product states~\cite{CPFSGBLPL10}. Another body of work has considered the setting in which the goal is not to learn the entire unknown quantum state $\rho$ but to learn only certain properties of $\rho$. In this context, people have considered tasks such as $(i)$ PAC learning: the learning algorithm here is given access to $(E_i,\Tr(\rho E_i))$ where  $\{E_i,\mathbb{I}-E_i\}$ is a uniformly random \gls{povm} element, $(ii)$ Shadow tomography, where the goal is, given copies of an unknown quantum state $\rho$, can we learn the expectation values of $\rho$ with respect to a certain set of \emph{fixed, a priori known} observables $\{E_1,\ldots,E_m\}$, $(iii)$ Other models such as classical shadows, online learning, learning with differential privacy that have modified the models $(i,ii)$. In all these models of learning, it is well-established  that~\cite{aaronson:shadow,aaronson2018online,aaronson2007learnability} the complexity of learning is $\mathcal{O}(n)$, which is \emph{exponentially} better than tomography.  For a detailed survey on the complexity of learning quantum data, we refer the interested reader to~\cite{anshusurvey}.

More practical applications of quantum data learning to \gls{hep} is to use it for extracting physical information from quantum states in quantum simulation.
This was first proposed in the context of condensed-matter physics~\cite{2019NatPh..15.1273C}
and further explored in~\cite{2021arXiv210612627H,2021PRXQ....2d0321B,PhysRevA.102.012415,2021arXiv210903400S,bernien2017probing, 2021arXiv210306712B,monaco2023quantum}.
The typical example is a recognition of quantum phases, where the \gls{qml} model learns the pair of quantum states and their phases to predict phase of unknown states.
In the context of high-energy physics, we often encounter phase transitions that cannot be investigated by local order parameters, such as confinement/deconfinement transition in \gls{qcd}.
It would be interesting to apply quantum data learning method to extract physical information in such situations.

In the longer-term, one may perform quantum experiment, not only digital quantum simulation but also analogue quantum simulation or others, then measure the final states via quantum sensor, and transduce the states coherently to a quantum computer which performs \gls{qml} to extract physical information~(see e.g., \cite{Huang_2022}).
This hybrid system could be extended to the concept of quantum-enhanced \gls{hep} experiment. A fascinating direction to exploit quantum data is to physically place quantum sensing devices in experiments and directly feed quantum states registered on the sensors into quantum computers. This certainly involves many challenges, e.g., detect particles or wave-like matters in quantum sensor, coherently transfer the generated state to other quantum systems, perform quantum operations to measure physical properties within coherence time. Such experiments will, however, provide an exciting opportunity to directly explore quantum phenomena observed in \gls{hep} experiments and extract dynamical properties of entangled quantum states.

A `\textit{conditio sine qua non}' for the success of this program in the era of noisy, near-term quantum devices is the co-design of error mitigation schemes that can efficiently compensate for  the different noise sources (e.g., gate errors, qubit decoherence and cross-talk) and guarantee results of sufficient quality to extract the physics of interest. To this end, several error mitigation schemes have been proposed in the past few years (see Sec.~\ref{sec:ibm_roadmap}) including zero-noise extrapolation~\cite{Temme2017Error}, probabilistic error cancellation~\cite{Berg2022Probabilistic}, and the probabilistic error amplification approach recently applied to the dynamics of the transverse field Ising model with more than 100 sites~\cite{Eddins_2023}.
All these methods will require an accurate description of all noise sources of current devices, which in the mean-time became a very active and successful area of research~\cite{
Bennett1996Purification, 
Kern2005Quantum, 
geller2013efficient, 
Temme2017Error, 
Kandala2019Error, 
Berg2022Probabilistic}. 
Finally, the precision of most quantum algorithms will depend on the quality of the measurement process of the observables of interest. Accurate results can require a number of projective measurements that can easily exceed what is currently affordable with the present gate times (from about hundred $ns$ with superconducting qubits, up to a few hundred $ms$ with ion-based technologies), which determine the clock-speed of quantum computing hardware calculations. 
Also in this case, there is the urge to design novel approaches capable of reducing the measurement overhead. 
Informationally complete \gls{povm}~\cite{PRXQuantum.2.040342} as well as classical shadows~\cite{Huang2020} offer viable solutions to this problem, opening new avenues for the use of quantum computing in large scale simulations.

\begin{acknowledgments}

A.D.M., M.G, and S.V.\ are supported by CERN through the CERN Quantum Technology Initiative (CERN QTI).
K.J.'s work is funded by the European Union’s Horizon Europe Framework Programme (HORIZON) under the ERA Chair scheme with grant agreement no.\ 101087126.
K.J., A.C., C.T., and S.K.\ are supported with funds from the Ministry of Science, Research and Culture of the State of Brandenburg within the Centre for Quantum Technologies and Applications (CQTA). 
\begin{center}
    \includegraphics[width=0.1\textwidth]{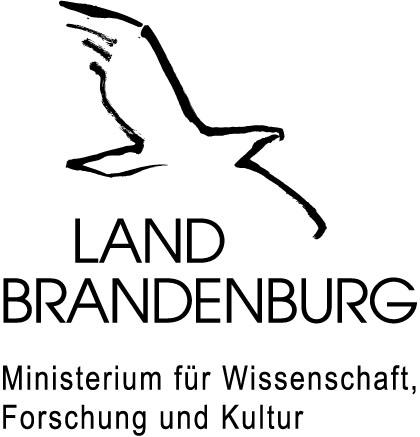}
\end{center}

A.R.\ is funded by the European Union. Views and opinions expressed are however those of the author(s) only and do not necessarily reflect those of the European Union or the European Commission. Neither the European Union nor the granting authority can be held responsible for them. This project has received funding from the European Union’s Horizon Europe research and innovation programme under grant agreement No.\ 101080086 NeQST.
\begin{center}
    \includegraphics[width=0.1\textwidth]{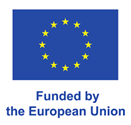}
\end{center}

E.R.O. is supported by the grant PID2021-126273NB-I00 and by the EU via QuantERA project T-NiSQ grant PCI2022-132984, QuantERA project QuantHEP, project Euryqa and PASQUANS2 funded by the European Union ``NextGenerationEU''/PRTR, by "ERDF A way of making Europe", by MCIN/AEI/10.13039/501100011033, and the Italian National center for HPC, Big Data and Quantum Computing, and the Basque Government through Grant No.\ IT1470-22. 

J.T.B.\ has received support from the European Union’s Horizon Europe research and innovation programme through the ERC StG FINE-TEA-SQUAD (Grant No. 101040729).
J.T.B., V.D. and V.C.\ are supported by the Dutch National Growth Fund (NGF), as part of the Quantum Delta NL programme.
V.D. and V.C also are supported by the Netherlands Organisation for Scientific Research (NWO/OCW), as part of the Quantum Software Consortium programme (project number 024.003.037 / 3368). ZH acknowledges support from the Sandoz Family Foundation-Monique de Meuron.

L. N.\ is supported by the IBM-UTokyo lab under the Japan-IBM Quantum Partnership.

J.S. would like to thank the support of Funda\c{c}\~ao para a Ci\^encia e a Tecnologia (FCT) under contracts CERN/FIS-COM/0036/2019 and UIDB/04540/2020 and project QuantHEP supported by the EU H2020 QuantERA ERA-NET Cofund in Quantum Technologies and by FCT (contract QuantERA/0001/2019).

E.F. acknowledges support by the Deutsche Forschungsgemeinschaft (DFG, German Research Foundation) under Germany’s Excellence Strategy –– EXC-2123 “QuantumFrontiers” — 390837967.

The IBM team acknowledges Jay Gambetta for his precious and constant support of the HEP working group. 
IBM, the IBM logo, and ibm.com are trademarks of International Business Machines Corp., registered in many jurisdictions worldwide. Other product and service names might be trademarks of IBM or other companies. The current list of IBM trademarks is available at \url{https://www.ibm.com/legal/copytrade}.

\end{acknowledgments}

\appendix
\section{Resource Requirements for Quantum Simulation of lattice \gls{qed} \label{appendix_resources}}

In this appendix we assess the resource requirements for the implementation of the quantum link model formulation of $U(1)$ lattice gauge theories with dynamical Wilson fermions in arbitrary dimension $d$. 
In reference~\cite{Mathis2020}, we assessed the number of qubits required to capture all degrees of freedom. Then we also reported the number of Pauli strings that is required to implement the different terms in the \gls{qed} Hamiltonian and finally, in the same publication we touched upon how this translates into the number of required quantum gates. 
We express all scalings in terms of a combination of the following model parameters:
\begin{itemize}
    \setlength\itemsep{0pt}
    \item \makebox[1.8cm]{$n_s$\hfill} number of lattice sites.
    \item \makebox[1.8cm]{$n_e$\hfill} number of lattice edges. Scales linearly with $n_s$ in regular lattices.
    \item \makebox[1.8cm]{$n_p$\hfill} number of lattice plaquettes. Scales linearly with $n_s$ in regular lattices.
    \item \makebox[1.8cm]{$n_\text{spinor}$\hfill} number of spinor components.
    \item \makebox[1.8cm]{$d$\hfill} number of spatial lattice dimensions.
    \item \makebox[1.8cm]{$d_S$\hfill} dimension of the spin $S$ system in the quantum link model.
    \item \makebox[1.8cm]{$n_\text{nonzero}(A)$\hfill} number of nonzero elements of the matrix $A$.
    \item \makebox[1.8cm]{$n_\text{pauli}(\hat{O})$\hfill} total number of Pauli strings in the encoding of the operator $\hat{O}$.
    \item \makebox[1.8cm]{$n_\text{real}(\hat{O})$\hfill} number of Pauli strings with real coefficients in the encoding of $\hat{O}$.
    \item \makebox[1.8cm]{$n_\text{imag}(\hat{O})$\hfill} number of Pauli strings with imaginary coefficients in the encoding of $\hat{O}$.
    \item \makebox[1.8cm]{$n_\text{mix}(\hat{O})$\hfill} number of Pauli strings with neither purely real nor purely imaginary coefficients 
    in the expansion of $\hat{O}$.
\end{itemize}

Table~\ref{4:tbl:hamiltonian_scaling} provides a summary
of this analysis. For the exact formulas for the number of Pauli terms, we refer to the respective sections above. 

\begin{table}[htb]
\centering
\begin{tabular}{@{}lccc@{}}\toprule
Term & \multicolumn{3}{c}{Number of Pauli strings} \\ 
\cmidrule{2-4} 
& $\log$. enc. & $\log$. enc. (perfect) & lin. enc. \\
\midrule

$H_\text{mass}$ & 
$\order{n_\text{s} n_\text{spinor}}$ & 
$\order{n_\text{s} n_\text{spinor}}$ & 
$\order{n_\text{s} n_\text{spinor}}$
\\

$H_\text{hopp}$ & 
$\order{n_\text{s} d n_\text{spinor}^2 d_S^2}$ &
$\order{n_\text{s} d n_\text{spinor}^2 d_S}$ &
$\order{n_\text{s} d n_\text{spinor}^2 d_S}$\\

$H_\text{wilson}$  & 
$\order{n_\text{s} d n_\text{spinor}^2 d_S^2}$ &
$\order{n_\text{s} d n_\text{spinor}^2 d_S}$ &
$\order{n_\text{s} d n_\text{spinor}^2 d_S}$
\\

$H_\text{elec}$ & 
$\order{n_s d d_S} $ &
$\order{n_s d d_S} $ &
$\order{n_\text{s} d d_S^2}$
\\

$H_\text{plaq}$ 
& $\order{n_s d d_S^8}$ &  $\order{n_s d d_S^4}$
& $\order{n_s d d_S^4}$
\\

\bottomrule
\end{tabular}
\centering
\label{4:tbl:hamiltonian_scaling}
\caption{The scaling relations for the number of Pauli terms for the terms in the lattice \gls{qed} Hamiltonian are shown for different encodings of the truncated gauge operators. These relations do not depend on whether the Jordan-Wigner, Bravyi-Kitaev or Parity mapping is used for the fermions.}
\end{table}

The analysis shows that the dominant term with respect to the number
of required Pauli strings is the plaquette term $H_\text{plaq}$, due to its strong scaling of with $d_S$. For this reason, even for small values of $d_S$ the plaquette term contributes by far the highest number of Pauli strings of all the terms in the Hamiltonian. 

The best overall scaling, and thus the lowest number of required Pauli terms, is achieved by using a logarithmic encoding for a quantum link model with a perfectly representable spin $S$ system. The logarithmic encoding is also more favorable in terms of the number of required qubits. The downside of perfectly representable $S$ is that the eigenvalue $S_z = 0$ is not contained in the spectrum as $d_S$ is a power of two, resulting in a degenerate ground state.

\section{Algorithms and Their Limitations\label{app:algos_limits}}

In this section we provide an overview over various classical and quantum algorithms relevant for the field of high-energy physics and highlight their capabilities as well as their limitations.

\subsection{VQE and VQD}

The variational quantum eigensolver is a hybrid quantum-classical approach to obtain an approximation for the ground state of a (quantum) system~\cite{Peruzzo2014}. The algorithm uses the quantum device to prepare an ansatz state in form of a parametric quantum circuit. Based on the measurement outcome of the expectation value of the Hamiltonian, a classical minimization algorithm is used to obtain a new set of parameters. Running the feedback loop between the classical computer and the quantum device until convergence, one obtains an approximation for the ground state and its energy, provided the chosen ansatz is expressive enough and the optimization did not converge to a local minimum. Main limitations of the \acrshort{vqe} are barren plateaus (see Sec.~\ref{subsec:qml_limitations}). 

\gls{vqd} is an extension of the \gls{vqe} allowing for computing low-lying excitations by running a \gls{vqe} looking for a low energy state that is orthogonal to all previous states~\cite{Higgott2019}.  
\gls{ssvqe} is another approach used to compute excited states. This algorithm searches a low energy
subspace by supplying orthogonal input states to the variational ansatz~\cite{PhysRevResearch.1.033062}. All the variational algorithms can be applied to Hamiltonians in both theoretical models and experimental analysis.

\subsection{Tensor Networks}

Tensor Networks are a family of entanglement-based ansätze providing an efficient parametrization of the physically relevant moderately entangled states~\cite{Banuls2019SimulatingLG,Banuls2020TNreview}. TN algorithms allow for computing ground states, low-lying excitations, thermal states and to a certain extent real-time dynamics. While TN are extremely successful in situations with moderate entanglement, they cease to work for highly entangled scenarios such as out-of-equilibrium dynamics. Moreover, in higher dimensions the numerical algorithms are computationally challenging but have a polynomial scaling in tensor size, thus allowed for first proof-of-principle demonstrations for \acrshort{lgt}s~\cite{Felser2019,Magnifico2020}.

\subsection{QAOA}

The quantum approximate optimization algorithm is a hybrid quantum-classical approach, originally designed to tackle combinatorial optimization problems~\cite{farhi2014quantum}. The problem is encoded in an Ising type Hamiltonian whose ground state is the optimal solution to the combinatorial optimization problem. \gls{qaoa} can be seen as a special type of \gls{vqe}, where the intial state is given by $\bigotimes \ket{+}$ and the parametric ansatz circuit in its plain vanilla form consists of a series of two alternating types of layers, each one containing a single real parameter. The first one is the exponential of the problem Hamiltonian, $\exp(-i\gamma \mathcal{H})$,  followed by a mixing layer corresponding to $R_X(\beta_i)$ gates applied to each qubit. In the limit of infinitely many layers, \gls{qaoa} can be interpreted as an adiabatic evolution of an eigenstate of the $X$ operator to the one of the problem Hamiltonian. From a theoretical point of view the performance of \gls{qaoa} is not entirely clear, it seems to depend on various factors and does not necessarily outperform classical algorithms~\cite{farhi2016quantum,barak_et_al:LIPIcs.ITCS.2022.14,Akshay2020}. 
Furthermore, the resulting quantum circuits can be deep making them hard to implement on noisy hardware~\cite{franca2020limitations, Weidenfeller2022scalingofquantum}. However, some of these issues may be alleviated by algorithmic advances such as warm-starts~\cite{Egger2021warmstartingquantum} and counteradiabatic driving~\cite{Wurtz2022counterdiabaticity}.

\subsection{QKMEANS}

The classical $k-$means is an efficient algorithm to classify data into $k$ clusters based on an unlabeled set of training vectors. 
It belongs to the family of unsupervised machine learning algorithms. The number $k$ of clusters must be known \textit{a priori} which somewhat limits the range of its application in HEP. The algorithm is iterative and assigns at each step a training vector to the nearest centroid. The centroid location is then updated according to the average over the cluster of vectors associated at the current step to the centroid. The most time/resources consuming part of the algorithm is the calculation of the distance.
In the classical version, using Lloyd's version of the algorithm, the time complexity is $\mathcal{O}(NM)$ where $N$ is the number of features and $M$ is the number of training examples~\cite{Lloyd82,Lloyd2013,Kopzyk2018}. The quantum version of the $k-$means algorithm provides an exponential speed-up for very large dimensions of a training vector. This is achieved through the introduction of two quantum subroutines, \textit{SwapTest} and \textit{DistCalc}, for the distance calculation~\cite{Aimeretal2006} and quantum subroutine \textit{GroverOptim} to assign a vector to the closest centroid cluster~\cite{Hoyer-Durr96}.

\subsection{Quantum Kernels}

Quantum kernels are a supervised quantum machine learning algorithm for classification and regression. The inputs can either be quantum (i.e., quantum states with an associated classical label) or fully classical (i.e. input-output data pairs). For the latter, the input classical data is first embedded into quantum states. For a quantum speed-up over classical algorithms, it is important to use an embedding (also called a quantum feature map) that is capable of recognizing classically intractable features~\cite{huang2021power,kubler2021inductive,liu2021rigorous}.
For a given input pair of inputs one then evaluates a similarity measure between two encoded quantum states on a quantum computer. Formally, this is function corresponds to an inner product of data states, and is known as a quantum kernel~\cite{schuld2021supervised,havlivcek2019supervised, huang2021power}. The fidelity quantum kernel~\cite{schuld2021supervised,havlivcek2019supervised} and projected quantum kernel~\cite{huang2021power} are two common choices in kernels.

\subsection{Quantum Generative Modelling}

Quantum systems, as inherently probabilistic systems, are naturally tailored to generative modelling tasks~\cite{PerdomoOrtiz2017}. The aim of generative modelling is to use training samples from a given target distribution to learn a model distribution which can then be used to generate new samples. As well as providing an efficient means of generating samples, it has been shown that quantum generative models can encode probability distributions that cannot be modelled efficiently classically~\cite{Coyle2019, sweke2020learnability, gao2021enhancing}. A number of different architectures and training strategies are being explored for quantum generative modelling. 
The Quantum circuit Born machine (QCBM)~\cite{benedetti2019generative} encodes a probability distribution in an $n$-qubit pure state. The Quantum Boltzmann Machine (QBM)~\cite{QBM_amin} is based on the Boltzmann distribution of a quantum Hamiltonian. A Quantum Generative Adversarial Network (QGAN)~\cite{QGAN_Loyd} uses the interplay of a generative quantum neural network and a classical or quantum discriminative model to a target distribution. 
In all cases the quantum generative model is generally trained by optimizing a cost function which estimates the distance between the model distribution and the training distribution. Commonly used costs include the KL divergence~\cite{kullback1951KLD}, the Jensen-Shannon divergence~\cite{lin1991divergence}, the (quantum) Rényi divergence~\cite{renyi1961measures, kieferova2021quantum} and the Maximum Mean Discrepancy~\cite{Gretton2012mmd}.

\subsection{QRL}	

\gls{rl} is an interactive mode of machine learning well suited for sequential decision and control tasks, and its objective is identifying the optimal policy (specification of what a learner does in a given situation) for a task environment. Current state-of-art methods include policy gradient methods, where the optimal policy is parametrized, and the performance is optimized in the policy space using interactions with the task environment; deep Q-learning methods, where the optimal value functions, which evaluate the ``value'' of a given state-action pair under a given policy, are approximated. Other approaches combine features of policy- and value-function-based methods.
In \gls{qrl}; i.e., in quantum approaches to \gls{rl}, the policies (in policy gradients), or value functions (in value-function-based methods) are expressed using parametrized quantum circuits, instead of, e.g., neural networks which are conventionally used.
The first quantum policy methods which achieved successful performances in OpenAI gym benchmarking environments were reported in~\cite{JerbiNEURIPS2021}, and the same paper proved the existence of task environments which can only be learned with quantum learners. In~\cite{SkolikQuantum2022} the quantum approach was extended to value-based approaches (deep Q-learning), and analogous proofs of learning separations were given.
The work~\cite{JerbiPRXQ2021} studies using quantum methods to speed up neural network-based deep energy models.
Follow-up works include the analysis of the performance of simple unentangled quantum learners~\cite{hsiao2022unentangled}, learning in partially observable environments~\cite{chen2022quantum}, applications in combinatorial optimization~\cite{skolik2022equivariant} and others. 
\gls{qrl} is also adopted in~\cite{schenk2022hybrid} where free energy-based reinforcement learning (FERL) is extended to multi-dimensional continuous state-action space environments to open the doors for a broader range of real-world applications.
An hybrid actor-critic scheme for continuous state-action spaces is developed based on the Deep Deterministic Policy Gradient algorithm combining a classical actor network with a QBM-based critic. The environments used throughout represent existing particle accelerator beam line of the Advanced Plasma Wakefield Experiment (AWAKE) at CERN.
\gls{qrl} with parameterized circuits suffers from barren plateaus as well (as it contains conventional supervised learning as a special case), although it is not known whether the phenomenon is exacerbated. In a recent work~\cite{skolik2022robustness}, the effect of noise was studied as well, and the results suggest the models could be somewhat resistant to noise, but more studies are required for conclusive findings.

\subsection{Topological Data Analysis}
\gls{tda} is an increasingly studied technique for extracting robust topological features from complex datasets and has in recent times also been employed in high-energy physics problems~\cite{sale2022probing}. The principal computational task in \gls{tda} is the extraction of so-called (persistent) Betti numbers, which can be used to distinguish the underlying topological spaces of data.  
 In the work~\cite{LloydNatComms2016}, a quantum algorithm for this problem was proposed, and it was suggested it may offer exponential speed-ups over conventional methods.
 In~\cite{GyurikQuantum2022, cade2021complexity} it was proven that certain generalizations of the \gls{tda} problem are DQC1-hard (and thus likely offer exponential speed-ups, and~\cite{HayakawaQuantum2022} showcases how persistent features can be extracted as well.
 The papers~\cite{ubaru2021quantum, berry2022quantifying, mcardle2022streamlined} provide streamlined versions of the original algorithm and achieve up-to-exponential savings in the qubit numbers, and~\cite{berry2022quantifying} has showcased a concrete family of datasets where concrete superpolynomial speed-ups over the best conventional methods are achieved.
 In~\cite{cade2021complexity, crichigno2022clique}, based on~\cite{WittenJDG1982}, a deep connection between TDA and supersymmetric theories has been established which may lead to new applications of (quantum) \gls{tda} in not only analysing experimental data, but also exploring theoretical spaces beyond the standard model.
 However it is important to note that it still remains to be determined if quantum \gls{tda} offers guaranteed speed-ups, or if it can be ``de-quantized'' using a new class of classical methods. Further, it is still an open question whether the regimes where quantum dramatic speed-ups kick in (i.e. when the desired homology, or Betti number, is high) have a wide application.

%

\end{document}